\documentclass[a4paper,11pt]{article}

\usepackage{jheppub}

\usepackage[utf8]{inputenc}
\usepackage[american]{babel}

\usepackage{comment}
\usepackage{fancyvrb} 

\usepackage{soul}
\usepackage{mathrsfs}
\usepackage{booktabs}

\usepackage{hyperref}
\hypersetup{%
    colorlinks,%
    breaklinks,%
    citecolor=blue,urlcolor=blue,linkcolor=blue,%
    hypertexnames=false,%
    pdftitle={The elliptic three-loop integrals of HVP in ChPT},%
    pdfauthor={L. Lellouch, A. Lupo, M. Sjo, P. Vanhove}%
}
\makeatletter
\renewcommand\HyPsd@CatcodeWarning[1]{}     
\makeatother

\usepackage{cleveref}
\crefname{equation}{eq.}{eqs.}
\crefname{section}{sec.}{secs.}
\Crefname{equation}{Eq.}{Eqs.}
\Crefname{section}{Sec.}{Secs.}
\newcommand{\rcite}[2][]{ref.~\cite[#1]{#2}}
\newcommand{\rrcite}[2][]{refs.~\cite[#1]{#2}}

\usepackage{afterpage}

\let\Re\relax\let\Im\relax
\DeclareMathOperator{\Re}{Re}
\DeclareMathOperator{\Im}{Im}

\DeclareMathOperator{\abs}{abs}

\usepackage{subcaption}
\usepackage{pgf,tikz}
\usetikzlibrary{%
    calc,
    positioning,
    patterns,
    decorations.pathreplacing,
    decorations.markings,
    decorations.pathmorphing,
    decorations.shapes,
    shapes.geometric,
    arrows.meta,
    external%
}

\usepackage{pgfplots}
\pgfplotsset{compat=1.18}
\usepgfplotslibrary{fillbetween}

\usepackage{mathtools} 
\usepackage{xfrac} 

\newcommand{\eqnbreak}{\notag\\&\quad}

\renewcommand{\d}{\mathrm{d}}
\newcommand{\p}{\partial}
\newcommand{\Z}{\mathbb{Z}}
\renewcommand{\O}{\mathcal{O}}

\newcommand{\chpt}{\ifmmode{\mathrm{ChPT}}\else{ChPT}\fi}
\newcommand{\LO}{\ifmmode{\mathrm{LO}}\else{LO}\fi}
\newcommand{\NLO}{\ifmmode{\mathrm{NLO}}\else{NLO}\fi}
\newcommand{\NNLO}{\ifmmode{\mathrm{NNLO}}\else{NNLO}\fi}
\newcommand{\NNNLO}{\ifmmode{\mathrm{N^3LO}}\else{N\textsuperscript{3}LO}\fi}

\DeclareMathOperator{\Li}{Li}

\DeclareMathOperator{\eLi}{\mathcal Li}
\DeclareMathOperator{\E}{\mathcal E}

\newcommand{\I}{\mathcal I}
\renewcommand{\IJ}[1]{\I_J^{(#1)}}
\newcommand{\IE}[1]{\I_E^{(#1)}}
\newcommand{\Irat}[1]{\I_\rat^{(#1)}}
\newcommand{\Idiv}{\I_\Div}
    
\newcommand{\Mpi}{M}

\newcommand{\Lpi}{L}

\newcommand{\rat}{\mathrm{rat}}
\newcommand{\Reg}{\mathrm{reg}}
\newcommand{\Div}{\mathrm{div}}

\newcommand{\Jbub}[1]{J_\bub^{(#1)}}

\newcommand{\Hbball}[1]{\mathcal{H}_\bball^{(#1)}}
\newcommand{\Dq}[2][]{\mathcal D_{\hspace{-.5ex}q}^{#1}[#2]}

\newcommand{\hypergeometric}[5][]{{}_{2\hspace{-1pt}}\mathit{F}_{\hspace{-1pt}1}#1(\genfrac{}{}{0pt}{}{#2,#3}{#4}#1|#5#1)}

\newcommand{\transf}[1]{\overset{#1}{\longrightarrow}}

\DeclareRobustCommand{\map}[2]{\langle#1 \mapsto #2\rangle} 

\newcommand{\bulletpoint}{\medskip\noindent $\bullet$ }
\newcommand{\logpm}[1]{\log\bigg(\frac{#1+1}{#1-1}\bigg)}
\newcommand{\logsq}[1]{\log\bigg(\frac{{#1}^2-1}{{#1}^2}\bigg)}

\definecolor{bubblegum}{rgb}{0.99, 0.76, 0.8}

\newcommand{\literedrefs}{\cite{Lee:2012cn,Lee:2013mka}}
\newcommand{\psdrefs}{\cite{Borowka:2017idc,Borowka:2018goh,Heinrich:2021dbf,Heinrich:2023til}}
\newcommand{\amfrefs}{\cite{Liu:2022chg}}
\newcommand{\ftrefs}{\cite{Borinsky:2020rqs,Borinsky:2023jdv}}
\newcommand{\mainref}[1][]{\cite[#1]{Lellouch:2025rnz}}
\newcommand{\rmainref}[1][]{\rcite[#1]{Lellouch:2025rnz}}

\newcommand{\tadpoleangle}{130}
\newlength{\tadpolesize}
\newlength{\bubblesize}
\newcommand{\bubbleaspect}{.8}
\newlength{\photonlength}
\newcommand{\diagramscale}{1.25}
\newlength{\vertsize}
\tikzset{
    Ultra thick/.style={line width=2pt},
    miter cap/.style={{Triangle Cap[]}-{Triangle Cap[]}},
    pion/.style = { very thick },
    photon/.style = { thick, decorate, decoration={snake,segment length=.26\photonlength, amplitude=+.05\photonlength} },
    notoph/.style = { thick, decorate, decoration={snake,segment length=.26\photonlength, amplitude=-.05\photonlength} },
    diagrcut/.style={red,decorate,decoration={zigzag,segment length=1mm,amplitude=.25mm}},
    bubble2/.style = {xscale=.8, transform shape},
    elbbub2/.style = {xscale=1.25, transform shape},
    bubble3/.style = {xscale=.64, transform shape},
    threeloop/.style = {scale=1.2, transform shape},
    NLO/.pic = {
        \fill (0,0) circle[radius=\vertsize];
    },
    NNLO/.pic = {
        \fill (-\vertsize,-\vertsize) rectangle (\vertsize, \vertsize);
    },
    NNNLO/.pic = {
        \fill (90:2\vertsize) -- (210:2\vertsize) -- (330:2\vertsize) -- cycle;
    },
    pics/tadpole/.style = {
        code = {
            \draw[pion] (0,0) .. controls (\tadpoleangle:\tadpolesize) and (180-\tadpoleangle:\tadpolesize) .. (0,0)
                foreach \i in {#1} { node[pos=.\i] (node\i) {} }
                ;
        }
    },
    pics/bubble/.style = {
        code = {
            \draw[pion] (0,0) ellipse [x radius=\bubblesize, y radius=\bubbleaspect\bubblesize];
            \foreach \i in {#1} {
                \coordinate (node\i) at
                (canvas polar cs:angle=\i, x radius=\bubblesize, y radius=\bubbleaspect\bubblesize) {};
            }
        }
    },
    pics/bubble/.default = {0}, 
    bubblejoint/.pic = {
        \fill[white] (-\bubblesize,-\bubblesize) rectangle (\bubblesize,\bubblesize);
        \draw[pion] (-\bubblesize,+\bubbleaspect\bubblesize) .. controls (0,+\bubbleaspect\bubblesize) and (0,-\bubbleaspect\bubblesize) .. (+\bubblesize,-\bubbleaspect\bubblesize);
        \draw[pion] (+\bubblesize,+\bubbleaspect\bubblesize) .. controls (0,+\bubbleaspect\bubblesize) and (0,-\bubbleaspect\bubblesize) .. (-\bubblesize,-\bubbleaspect\bubblesize);
    },
}

\newcounter{diagram}

\definecolor{plotI}{HTML}{0077BB}
\definecolor{plotII}{HTML}{33BBEE}
\definecolor{plotIII}{HTML}{009988}
\definecolor{plotIV}{HTML}{EE7733}
\definecolor{plotV}{HTML}{CC3311}
\definecolor{plotVI}{HTML}{DDAA33}
\definecolor{plotVII}{HTML}{AA3377}

\colorlet{spacelike}{plotI}
\colorlet{subthr}{plotIV}
\colorlet{twopi}{plotIII}
\colorlet{fourpi}{plotVII}
\colorlet{shaded}{gray!30}

\usepackage{etoolbox}

\definecolor{plotI}{HTML}{0077BB}
\definecolor{plotII}{HTML}{33BBEE}
\definecolor{plotIII}{HTML}{009988}
\definecolor{plotIV}{HTML}{EE7733}
\definecolor{plotV}{HTML}{CC3311}
\definecolor{plotVI}{HTML}{DDAA33}
\definecolor{plotVII}{HTML}{AA3377}

\tikzset{
    dash0/.style={solid},
    dash1/.style={dash pattern=on 8pt off 2pt on 2pt off 2pt},
    dash2/.style={dash pattern=on 6pt off 2pt on 2pt off 2pt on 2pt off 2pt},
    dashA/.style={dash pattern=on 4.4pt off 4.4pt},
    dashB/.style={dash pattern=on 2pt off 2pt},
    dashC/.style={dash pattern=on 1.2pt off 1pt},
    dash3/.style={dash pattern=on 5.5pt off 1.5pt on 1.5pt off 1.5pt on 1.5pt off 1.5pt on 1.5pt off 1.5pt},
    dash4/.style={dash pattern=on 5.2pt off 1.2pt on 1.2pt off 1.2pt on 1.2pt off 1.2pt on 1.2pt off 1.2pt on 1.2pt off 1.2pt},
    dash5/.style={dash pattern=on 5.2pt off 0.9pt on 0.9pt off 0.9pt on 0.9pt off 0.9pt on 0.9pt off 0.9pt on 0.9pt off 0.9pt},
    line0/.style={plotI,   dash0},
    line1/.style={plotII,  dash1},
    line2/.style={plotIII, dash2},
    lineA/.style={plotIV,  dashA},
    lineB/.style={plotV,   dashB},
    line3/.style={plotVI,  dash3},
    line4/.style={plotVII, dash4},
    line5/.style={plotVII, dash5},
    real/.style={line0, thick},
    imag/.style={lineA, thick},
    bessel/.style={dashA, plotVII, thick},
    polylog/.style={dash2, plotVI, thick},
    eisen/.style={dash0, plotIII, thick},
    poisson/.style={dash1, plotIV, thick},
    series dash/.style={dashC, thick},
    series/.style={series dash, plotV},
    secdec/.style={dashB, plotI, thick, join=round},
    feyntrop/.style={dash4, plotII, thick},
    amflow/.style={solid, thick, mark=+, draw=none, /pgfplots/error bars/y dir=both, /pgfplots/error bars/y explicit},
    }

\colorlet{E1}{plotI}
\colorlet{E2}{plotIII}
\colorlet{E3}{plotV}
\colorlet{E4}{plotIV}
\colorlet{E5}{plotVI}
\colorlet{E6}{plotVII}
\colorlet{realextra}{plotIV}

\tikzset{
    branchcut/.style={decorate, decoration={zigzag, segment length=1mm, amplitude=.4mm}, rounded corners=.01pt},
    realpos/.style={thick, plotV},
    realneg/.style={thick, plotII},
    nocut/.style={densely dotted},
    C0/.style={-{>[scale=.7]}},
    C1/.style={{<[scale=.7]}-{>[scale=.7]}},
    realline/.style={line width=2pt, miter cap},
    realline shadow/.style={white, dashed, thick},
    }

\pgfplotsset{
    every axis/.style={set layers},
    tcontour/.style={very thin, gray, mark=none, domain=-100:100, restrict y to domain=#1}}

\tikzset{
    threshold/.style={dash pattern=on 0pt off 2pt, cap=round, line width=1.5pt, opacity=.6},
    real/.style={plotI, very thick, dash0},
    imag/.style={plotIV, very thick, dashA},
    ours/.style={real, very thick, solid},
    errorband/.style={solid, draw=none, opacity=.35},
    faint errorband/.style={opacity=.35},
    dark errorband/.style={opacity=.6},
    total/.style={line width=2pt, black}}

\newlength{\legendwidth}
\pgfplotsset{
    complex legend/.style={legend image code/.code={
        \draw [real, #1] (0pt,.5ex) -- (\legendwidth,.5ex);
        \draw [imag, #1] (0pt,-.2ex) -- (\legendwidth,-.2ex);
        }},
    }
\pgfplotsset{
    legend with error/.style={legend image code/.code={
        \fill [#1, solid, errorband] (0pt,+.7ex) rectangle (\legendwidth,-.3ex);
        \draw [#1]            (0pt,+.2ex) -- (\legendwidth,+.2ex);
        }},
    legend with different error/.style 2 args={legend image code/.code={
        \fill [#2, errorband] (0pt,+.7ex) rectangle (\legendwidth,-.3ex);
        \draw [#1]            (0pt,+.2ex) -- (\legendwidth,+.2ex);
        }},
    double legend/.style={legend image code/.code={
        \draw [ours,     #1] (0pt,+.7ex) -- (\legendwidth,+.7ex);
        \draw [secdec,   #1] (0pt,+.1ex) -- (\legendwidth,+.1ex);
        }},
    triple legend/.style={legend image code/.code={
        \draw [ours,     #1] (0pt,+.7ex) -- (\legendwidth,+.7ex);
        \draw [secdec,   #1] (0pt,+.1ex) -- (\legendwidth,+.1ex);
        \draw [feyntrop, #1] (0pt,-.5ex) -- (\legendwidth,-.5ex);
        }},
    }

\newcommand{\plotEsymlog}[3]{%
    \begin{tikzpicture}
        \def\Eplotfile{\plotdir/prec64/E#1}
        \def\Eplotlabel{#2}
        \def\Eplotthreshold{#3}
        \def\symlog{_symlog-15-1}
        \input{\plotdir/prec64/E#1e\symlog.tex}
        \begin{axis}[
            width=\Eplotwidth, height=\symlogEplotheight, scale only axis,
            xmin=\Eplotxmin, xmax=\Eplotxmax, xlabel={},
            xtick={0,4,16}, xmajorgrids=true, xticklabels={},
            minor xtick={-9,-8,...,23},
            ymin=\symlogymin, ymax=\symlogymax, ylabel={\small$\Re\Eplotlabel$},
            ytick=\symlogsparseytick,
            yticklabels={,$-10^{-6}$,,$-10^{-12}$,,$0$,,$+10^{-12}$,,$+10^{-6}$,},
            yticklabel style={font=\tiny},
            ymajorgrids=true, yminorgrids=true
            ]
            \addplot[gray, amflow] table[x=t, y=NormRe, y error plus expr=abs(\thisrow{NormMaxRe}-\thisrow{NormRe}), y error minus expr=abs(\thisrow{NormMinRe}-\thisrow{NormRe})] {\Eplotfile a\symlog.dat};

            \addplot[eisen] table[x=t, y=NormRe] {\Eplotfile e\symlog.dat};
            \addplot[draw=none, name path=upper] table[x=t, y=NormMaxRe] {\Eplotfile e\symlog.dat};
            \addplot[draw=none, name path=lower] table[x=t, y=NormMinRe] {\Eplotfile e\symlog.dat};
            \addplot[eisen, errorband] fill between[of=lower and upper];

            \addplot[secdec] table[x=t, y=NormRe] {\Eplotfile s\symlog.dat};
            \addplot[draw=none, name path=upper] table[x=t, y=NormMaxRe] {\Eplotfile s\symlog.dat};
            \addplot[draw=none, name path=lower] table[x=t, y=NormMinRe] {\Eplotfile s\symlog.dat};
            \addplot[secdec, errorband] fill between[of=lower and upper];
            \ifnumless{#1}{5}{%
                \begin{scope}[/pgfplots/restrict x to domain=\Eplotxmin:\Eplotthreshold]
                    \addplot[feyntrop] table[x=t, y=NormRe] {\Eplotfile f\symlog.dat};
                    \addplot[draw=none, name path=upper] table[x=t, y=NormMaxRe] {\Eplotfile f\symlog.dat};
                    \addplot[draw=none, name path=lower] table[x=t, y=NormMinRe] {\Eplotfile f\symlog.dat};
                    \addplot[feyntrop, errorband] fill between[of=lower and upper];%
                \end{scope}
                }{}

            \addplot[series, restrict x to domain=-\Eplotthreshold:\Eplotthreshold] table[x=t, y=NormRe] {\Eplotfile 0\symlog.dat};
            \addplot[draw=none, name path=upper, restrict x to domain=-\Eplotthreshold:\Eplotthreshold] table[x=t, y=NormMaxRe] {\Eplotfile 0\symlog.dat};
            \addplot[draw=none, name path=lower, restrict x to domain=-\Eplotthreshold:\Eplotthreshold] table[x=t, y=NormMinRe] {\Eplotfile 0\symlog.dat};
            \addplot[series, errorband] fill between[of=lower and upper];

        \end{axis}

        \begin{axis}[
            shift={(0,-(\symlogEplotheight + \Eplotsep))},
            width=\Eplotwidth, height=\symlogEplotheight, scale only axis,
            xmin=\Eplotxmin, xmax=\Eplotxmax, xlabel={\xlabel},
            xtick={0,4,16}, xmajorgrids=true,
            minor xtick={-9,-8,...,23},
            ymin=\symlogymin, ymax=\symlogymax, ylabel={\small$\Im\Eplotlabel$},
            ytick=\symlogsparseytick,
            yticklabels={,$-10^{-6}$,,$-10^{-12}$,,$0$,,$+10^{-12}$,,$+10^{-6}$,},
            yticklabel style={font=\tiny},
            ymajorgrids=true, yminorgrids=true,
            axis background/.style={%
                preaction={
                    path picture={
                        \path[pattern=north east lines, pattern color=gray!50] (axis cs:\Eplotxmin,\symlogymax) rectangle (axis cs:\Eplotthreshold,\symlogymin);
                    }}}
            ]
            \addplot[gray, amflow, restrict x to domain=\Eplotxmin:\Eplotthreshold] table[x=t, y=Im, y error plus expr=abs(\thisrow{MaxIm}-\thisrow{Im}), y error minus expr=abs(\thisrow{MinIm}-\thisrow{Im})] {\Eplotfile a\symlog.dat};
            \addplot[gray, amflow, restrict x to domain=\Eplotthreshold:\Eplotxmax] table[x=t, y=NormIm, y error plus expr=abs(\thisrow{NormMaxIm}-\thisrow{NormIm}), y error minus expr=abs(\thisrow{NormMinIm}-\thisrow{NormIm})] {\Eplotfile a\symlog.dat};

            \addplot[eisen, restrict x to domain=\Eplotxmin:\Eplotthreshold] table[x=t, y=Im] {\Eplotfile e\symlog.dat};
            \addplot[draw=none, name path=upper, restrict x to domain=\Eplotxmin:\Eplotthreshold] table[x=t, y=MaxIm] {\Eplotfile e\symlog.dat};
            \addplot[draw=none, name path=lower, restrict x to domain=\Eplotxmin:\Eplotthreshold] table[x=t, y=MinIm] {\Eplotfile e\symlog.dat};
            \addplot[eisen, errorband] fill between[of=lower and upper];

            \addplot[secdec, restrict x to domain=\Eplotxmin:\Eplotthreshold] table[x=t, y=Im] {\Eplotfile s\symlog.dat};
            \addplot[draw=none, name path=upper, restrict x to domain=\Eplotxmin:\Eplotthreshold] table[x=t, y=MaxIm] {\Eplotfile s\symlog.dat};
            \addplot[draw=none, name path=lower, restrict x to domain=\Eplotxmin:\Eplotthreshold] table[x=t, y=MinIm] {\Eplotfile s\symlog.dat};
            \addplot[secdec, errorband] fill between[of=lower and upper];
            \ifnumless{#1}{5}{%
                \addplot[feyntrop, restrict x to domain=\Eplotxmin:\Eplotthreshold] table[x=t, y=NormRe] {\Eplotfile f\symlog.dat};
                \addplot[draw=none, name path=upper, restrict x to domain=\Eplotxmin:\Eplotthreshold] table[x=t, y=NormMaxRe] {\Eplotfile f\symlog.dat};
                \addplot[draw=none, name path=lower, restrict x to domain=\Eplotxmin:\Eplotthreshold] table[x=t, y=NormMinRe] {\Eplotfile f\symlog.dat};
                \addplot[feyntrop, faint errorband] fill between[of=lower and upper];%
                }{}

            \addplot[eisen, restrict x to domain=\Eplotthreshold:\Eplotxmax] table[x=t, y=NormIm] {\Eplotfile e\symlog.dat};
            \addplot[draw=none, name path=upper, restrict x to domain=\Eplotthreshold:\Eplotxmax] table[x=t, y=NormMaxIm] {\Eplotfile e\symlog.dat};
            \addplot[draw=none, name path=lower, restrict x to domain=\Eplotthreshold:\Eplotxmax] table[x=t, y=NormMinIm] {\Eplotfile e\symlog.dat};
            \addplot[eisen, errorband] fill between[of=lower and upper];

            \addplot[secdec, restrict x to domain=\Eplotthreshold:\Eplotxmax] table[x=t, y=NormIm] {\Eplotfile s\symlog.dat};
            \addplot[draw=none, name path=upper, restrict x to domain=\Eplotthreshold:\Eplotxmax] table[x=t, y=NormMaxIm] {\Eplotfile s\symlog.dat};
            \addplot[draw=none, name path=lower, restrict x to domain=\Eplotthreshold:\Eplotxmax] table[x=t, y=NormMinIm] {\Eplotfile s\symlog.dat};
            \addplot[secdec, dark errorband] fill between[of=lower and upper];

        \end{axis}
        \useasboundingbox (0,-\symlogEplotheight) + (0,-\Eplotsep) + (0,-.5cm); 
    \end{tikzpicture}
    }

\newcommand{\plotTime}[1]{
    \def\plotfile{\plotdir/prec64/E#1}
    \ifnumless{#1}{5}{%
        \addplot[imag, ours  ,   E#1, triple legend={E#1}] table[x=t, y expr=\thisrow{Time}/1e9] {\plotfile e.dat};
        \addplot[imag, secdec,   E#1, forget plot        ] table[x=t, y expr=\thisrow{Time}/1e9] {\plotfile s.dat};
        \addplot[imag, feyntrop, E#1, forget plot        ] table[x=t, y expr=\thisrow{Time}/1e9] {\plotfile f.dat};
        }{%
        \addplot[imag, ours  ,   E#1, double legend={E#1}] table[x=t, y expr=\thisrow{Time}/1e9] {\plotfile e.dat};
        \addplot[imag, secdec,   E#1, forget plot        ] table[x=t, y expr=\thisrow{Time}/1e9] {\plotfile s.dat};
        }
        index = 0
    }

\newcommand{\plotJbub}[1]{%
    \addplot[real, E#1, complex legend={E#1}] table[x=t, y=ReJ#1, col sep=tab] {\plotdir/Jbub#1.dat};
    \addplot[imag, E#1, forget plot, restrict x to domain=4:\Eplotxmax] table[x=t, y=ImJ#1, col sep=tab] {\plotdir/Jbub#1.dat};
    }

\newcommand{\drawzerolines}[4]{
            \draw[thin, gray] (#1, 0) -- (#2, 0);
            \draw[thin, gray] (0, #3) -- (0, #4);
            }

\newcommand{\newsetlength}[2]{\newlength{#1}\setlength{#1}{#2}}
\newsetlength{\Eplotwidth}{6cm}
\newsetlength{\Eplotheight}{3cm}
\newsetlength{\Eplotsep}{1mm}
\newsetlength{\normEplotheight}{.8cm}
\newsetlength{\doublenormEplotheight}{2\normEplotheight}
\newsetlength{\symlogEplotheight}{.7\Eplotheight}

\newcommand{\Eplotxmin}{-8}
\newcommand{\Eplotxmax}{+24}
\newcommand{\Eplotlogymin}{1e-18}
\newcommand{\Eplotlogymax}{1}
\newcommand{\xlabel}{}

\newcommand{\samplelength}{23pt}
\newcommand{\sample}[1]{\tikz[baseline=-0.25ex]{
    \draw[real, secdec, #1] (0,0) -- (\samplelength,0);
    \draw[real, ours,   #1] (0,.5ex) -- (\samplelength,.5ex);
    }}
\newcommand{\Resample}{%
    blue, solid:%
    \tikzset{external/export next=false}
    \tikz[baseline=-0.5ex]{
        \draw[real] (0,0) -- (19.5pt,0);
        }}
\newcommand{\Imsample}{%
    orange, dashed:%
    \tikzset{external/export next=false}
    \tikz[baseline=-0.6ex]{
        \draw[imag] (0,0) -- (19.5pt,0);
        }}
\newcommand{\seriessample}{%
    blue, dotted:%
    \tikzset{external/export next=false}
    \tikz[baseline=-0.6ex]{
        \draw[real, series] (0,0) -- (\samplelength,0);
        }}

\newcommand{\linesample}[1]{%
    \tikzset{external/export next=false}
    \tikz[baseline=-0.6ex]{
        \draw[#1] (0,0) -- (19.5pt,0);
        }}

\newsetlength{\hybridplotwidth}{.8\Eplotwidth}
\newsetlength{\hybridplotheight}{\symlogEplotheight}

\tikzset{
    totval/.style={plotI, thick, solid},
    headval/.style={plotII, thick, densely dashed},
    tailval/.style={plotIII, thick, densely dotted},
    }

\pgfplotsset{
    discard if/.style 2 args={
        x filter/.code={
            \edef\tempa{\thisrow{#1}}
            \edef\tempb{#2}
            \ifx\tempa\tempb
                \def\pgfmathresult{inf}
            \fi
        }
    },
    discard if not/.style 2 args={
        x filter/.code={
            \edef\tempa{\thisrow{#1}}
            \edef\tempb{#2}
            \ifx\tempa\tempb
            \else
                \def\pgfmathresult{inf}
            \fi
        }
    }
}

\newcommand{\plothybrid}[9]{%
    \def\hybridfile{\plotdir/hybrid_err_t-2.0.dat}
    \def\plainname{#1}
    \def\prettyname{#2}
    \def\chimin{#3}
    \def\chiref{#4}
    \def\chimax{#5}
    \def\relmin{#6}
    \def\relmax{#7}
    \def\valatchiref{#8}
    \def\errmin{#9}
    \begin{tikzpicture}
        \begin{semilogxaxis}[
            width=\hybridplotwidth, height=\hybridplotheight, scale only axis,
            xmin=\chimin, xmax=\chimax,
            xticklabels={},
            ylabel={$\prettyname (\text{relative})$}
            ]
            \addplot[mark=*] coordinates { (\chiref, 1) };
            \addplot[totval] table[x=chi, y expr={\thisrow{result}/\valatchiref}, discard if not={source}{\plainname}, col sep=tab] {\hybridfile};

        \end{semilogxaxis}
        \begin{loglogaxis}[
            shift={(0,-(\hybridplotheight + \Eplotsep))},
            xmin=\chimin, xmax=\chimax,
            ymin=\errmin,
            width=\hybridplotwidth, height=\hybridplotheight, scale only axis,
            xlabel={$\chi$}, xticklabel pos=bottom,
            ylabel={$\delta\prettyname$}
            ]
            \addplot[tailval] table[x=chi, y=dtail, discard if not={source}{\plainname}, col sep=tab] {\hybridfile};
            \addplot[headval] table[x=chi, y=dhead, discard if not={source}{\plainname}, col sep=tab] {\hybridfile};
            \addplot[totval] table[x=chi, y=dresult, discard if not={source}{\plainname}, col sep=tab] {\hybridfile};
        \end{loglogaxis}

        \useasboundingbox (0,-1.2\hybridplotheight) + (0,-.8cm); 
    \end{tikzpicture}}

\newcommand{\scriptdiagrscale}{0.3}
\newlength{\scriptdiagrleglen}
\newlength{\scriptdiagrleg}
\newcommand{\tad}{%
    {\tikz[scale=\scriptdiagrscale, transform shape]{%
        \tikzset{pion/.style={ thin }}
        \path[use as bounding box] (-1.5\scriptdiagrleglen,0) rectangle (+1.5\scriptdiagrleglen,.6\tadpolesize);
        \draw[pion] (-1.5\scriptdiagrleglen,0) -- (0,0) pic {tadpole} -- (+1.5\scriptdiagrleglen,0);}}}
\newcommand{\bub}{%
    {\tikz[scale=\scriptdiagrscale, transform shape]{%
        \tikzset{pion/.style={ thin }}
        \draw (0,0) pic {bubble};
        \draw[pion] (0,0) + (-\bubblesize,0) -- +(-\scriptdiagrleg,0);
        \draw[pion] (0,0) + (+\bubblesize,0) -- +(+\scriptdiagrleg,0);
        }}}
\newcommand{\sunset}{%
    {\tikz[scale=\scriptdiagrscale, transform shape]{%
        \tikzset{pion/.style={ thin }}
        \renewcommand{\bubbleaspect}{1}
        \draw[pion] (0,0) -- (-\bubblesize,0) -- +(-\scriptdiagrleglen,0);
        \draw[pion] (0,0) -- (+\bubblesize,0) -- +(+\scriptdiagrleglen,0);
        \draw (0,0) pic {bubble};
        }}}
\newcommand{\bball}{%
    {\tikz[scale=\scriptdiagrscale, transform shape]{%
        \tikzset{pion/.style={ thin }}
        \renewcommand{\bubbleaspect}{1}
        \draw[pion] (0,0) + (-\bubblesize,0) -- +(-\scriptdiagrleg,0);
        \draw[pion] (0,0) + (+\bubblesize,0) -- +(+\scriptdiagrleg,0);
        \draw (0,0) pic {bubble} pic[yscale=.4] {bubble};
        }}}
\newcommand{\fatcat}{%
    {\tikz[scale=\scriptdiagrscale, transform shape]{%
        \tikzset{pion/.style={ thin }}
        \renewcommand{\bubbleaspect}{1}
        \draw[pion] (0,0) + (-\bubblesize,0) -- +(-\scriptdiagrleg,0);
        \draw[pion] (0,0) + (+\bubblesize,0) -- +(+\scriptdiagrleg,0);
        \draw (0,0) pic {bubble=40};
        \draw (node40) pic[rotate=-40, scale=.6] {bubble};
        }}}
\newcommand{\catseye}{%
    {\tikz[scale=\scriptdiagrscale, transform shape]{%
        \tikzset{pion/.style={ thin }}
        \renewcommand{\bubbleaspect}{1}
        \draw[pion] (0,0) + (-\bubblesize,0) -- +(-\scriptdiagrleg,0);
        \draw[pion] (0,0) + (+\bubblesize,0) -- +(+\scriptdiagrleg,0);
        \draw (0,0) pic {bubble} pic[xscale=.4] {bubble};
        }}}
        
\newcommand{\catseyeplus}{%
    {\tikz[scale=\scriptdiagrscale, transform shape]{%
        \tikzset{pion/.style={ thin }}
        \renewcommand{\bubbleaspect}{1}
        \draw[pion] (0,0) + (-\bubblesize,0) -- +(-\scriptdiagrleg,0);
        \draw[pion] (0,0) + (+\bubblesize,0) -- +(+\scriptdiagrleg,0);
        \draw (0,0) pic {bubble={60,120,270}};
        \draw[pion] (node60)  -- (node270);
        \draw[pion] (node120) -- (node270);
        }}}    
\newcommand{\catseyeplusplus}{%
    {\tikz[scale=\scriptdiagrscale, transform shape]{%
        \tikzset{pion/.style={ thin }}
        \renewcommand{\bubbleaspect}{1}
        \draw[pion] (0,0) + (-\bubblesize,0) -- +(-\scriptdiagrleg,0);
        \draw[pion] (0,0) + (+\bubblesize,0) -- +(+\scriptdiagrleg,0);
        \draw (0,0) pic {bubble={60,120,240,300}};
        \draw[pion] (node60)  to[bend right=30] (node300);
        \draw[pion] (node240) to[bend right=30] (node120);
        }}}

\newlength{\inlinetikzheight}
\newcommand{\inlinetikz}[2]{{%
    \setbox0=\hbox{#1}\tikzset{external/export next=false}%
    \setlength{\inlinetikzheight}{\dimexpr\ht0+\dp0\relax}%
    \tikz[baseline=\dp0]{#2}%
    }}

\newcommand{\hardwarefn}{\textsuperscript{\mbox{\scriptsize\ref{fn:hardware}}}}

\newcommand{\figellipticcaption}{%
    \caption[]{The integrals $E_1$, $E_2$, and $E_3$ (left to right), shown as functions of the nome $q$ over the complex unit disk.
    This disk corresponds to a fractal tiling of infinitely many copies of the complex $t$-plane (see \cref{fig:q}), and we outline one such copy in white here.
    For each integral, its value at $t=0$ (limit as $q\to1$) has been subtracted to make the features more visible.
    The integral values are shown using domain coloring: the hue represents their phase (with \textbf{\color{red}red} and \textbf{\color{cyan}cyan} for positive and negative real values, respectively), while the contours represent their absolute value. The latter is a power of $2$ on the contours, and decreases toward the shaded side.
    As a reference for reading the contours, the top corner of the white outline (corresponding to $t=4$) has \mbox{$0.5 < |E_1(2;4)-E_1(2;0)| < 1$}, \mbox{$0.25<|E_2(2;4)-E_2(2;0)|<0.5$}, and \mbox{$0.125<|E_3(2;4)-E_3(2;0)|< 0.25$}.
    Note the branch cut on $q\in[-1,0]$, which corresponds to two copies of $t\in[16,\infty)$ (see \cref{fig:q}).
    }}

\newcommand{\figtaucaption}{%
    \caption{\textbf{Left}: the mapping $\tau\mapsto t$ visualized in the complex plane, covering one period in the real direction.
    The complex values are represented as in \cref{fig:elliptic}.
    The $\times$-shaped contour intersections are saddle points at $t=4$ (for instance at \mbox{$\tau=\pm\frac14+\frac{i\sqrt3}{12}$}) and $t=16$ (for instance at \mbox{$\tau=\pm\frac12+\frac{i\sqrt3}{6}$}).
    \textbf{Right}: a sketch (to scale) of the $\Re(\tau)>0$ half of the same plot, showing a selection of the infinitely many $\tau$-plane images of the real $t$-axis, colored according to which kinematic region they reproduce: spacelike subthreshold (\mbox{$t<0$}, \textbf{\color{spacelike}blue}), timelike subthreshold (\mbox{$0<t<4$}, \textbf{\color{subthr}orange}), two-pion (\mbox{$4<t<16$}, \textbf{\color{twopi}green}) and multi-pion (\mbox{$16<t$}, \textbf{\color{fourpi}violet}).
    The shaded area (if extended upward to infinity) contains all $t$ with $\Im(t)>0$ exactly once, with contours indicating (bottom to top) $\Im(t)=0.001$, $0.1$, $0.5$, and $1$; its boundary (bold) contains all real $t$ exactly once, ordered counterlockwise.}}

\newcommand{\figqcaption}{%
    \caption{Like \cref{fig:tau}, but showing $q\mapsto t$ over the unit disk.
            The lines in the sketch have been extended onto the $\Re(\tau)<0$ part (lower half of the disk), and thereby produce the outline used in \cref{fig:elliptic}.
            Note how inside the shaded region, $q$ approaches $1$ extremely slowly as $|t|$ approaches $0$, ensuring good convergence of our sums almost everywhere.
        }}

\newcommand{\figtsuncaption}{%
    \caption[]{Like \cref{fig:tau}, but showing {\tikzexternaldisable $t_\sunset\mapsto t$} as defined in \cref{eq:t-to-tau}.
        In the sketch on the right, the shaded area is $\Omega_>$, and the two inverse images of the real $t$-line are shown with bold colored lines as in \cref{fig:tau}.
        The dotted rectangle indicates the region shown in \cref{fig:rho24} (bottom row) and \cref{fig:rhoF}.
        The thin gray lines are the same contours of constant $\Im(t)$ as in \cref{fig:tau}, and the dashed ones indicate their images under the involutory mapping $z\mapsto (9z-9)/(z-9)$, which is how they appear in $\varpi_r$.}}

\newcommand{\figrhotwofourcaption}{%
    \caption{
        The construction of $\rho_2(z)$ (top row) and $\rho_4(z)$ (bottom row).
        \textbf{Left}: The principal branch of $\sqrt{(z-4)(z-16)}$ and $\sqrt[4]{z^3 - 9z^2 + 3z - 3}$, respectively, visualized in the complex plane using colors and contours like in \cref{fig:tau}.
        \textbf{Middle}: The corrected functions $\rho_2(z)$ and $\rho_4(z)$, respectively.
        \textbf{Right}: A schematic picture of their relation.
        Colored lines indicate where the argument of the root [$(z-4)(z-16)$ and $z^3 - 9z^2 + 3z - 3$, respectively] is real: \textbf{\color{plotV} red} for positive values and \textbf{\color{plotII} cyan} for negative.
        Dotted lines do not cause branch cuts.
        Solid lines cause cuts that are removed by analytically continuing to different branches, starting on the principal branch in the region marked by $\star$ and picking up roots of unity ($-1$ and $+i$, respectively) when crossing a line in the direction indicated by the arrows.
        Zigzag lines are left as branch cuts also in $\rho_2(z)$ or $\rho_4(z)$.
        The dotted circle indicates the boundary between $\Omega_<$ and $\Omega_>$ (compare \cref{fig:tsun}).
        }}

\newcommand{\figrhoFcaption}{%
    \caption[The function $\rho_F(z)-F_1(1)$, visualized over the same region as $\rho_4(z)$ in \cref{fig:rho24}.]{
        \textbf{Top}:
        The function $\rho_F(z)-F_1(1)$, visualized over the same region as $\rho_4(z)$ in \cref{fig:rho24}.
        Subtracting the constant $F_1(1)$ makes the features more visible.
        The upper half-plane shows $\rho_F(z)$ as used for the
        $t\mapsto\tau$
        relation, and the lower shows the uncorrected $F_1\big(1728/j_\sunset(z)\big)$ using the principal branch of the hypergeometric function.
        Both are symmetric (up to phases) across the real axis.
        The insets show the region around $z=9$, which for the uncorrected function contains the only discontinuity outside $\Omega_<$, making its removal the most important part of the construction.
        \textbf{Bottom}:
        A schematic picture of the construction of the correct $\rho_F(z)$,
        using the same line conventions as in \cref{fig:rho24}: dotted for no cut, solid for removed cut, zigzag for remaining cut.
        Negative real $z$ values only produce a cut for $F_2(z)$ and therefore do not affect the principal branch ($\star$), which is purely $F_1(z)$.
        Positive real $z$ values only produce a cut when \mbox{$|1728/j_\sunset(z)|\geq 1$}, which holds within the \textbf{\color{realextra} orange} area.
        The monodromies described in the text are used when crossing the solid lines to keep $\rho_F(z)$ continuous: $C_0$ ($\rightarrow$) or $C_1$ ($\leftrightarrow$) is applied when going in the direction of the arrows (recall that $C_1$ is involutory, hence the bidirectional arrow).
        }}

\newcommand{\figrhoFconstrcaption}{%
    \caption[ A schematic version of \cref{fig:rhoF}, showing the construction of the correct $\rho_F(z)$.]{
        A schematic version of \cref{fig:rhoF}, showing the construction of the correct $\rho_F(z)$.
        The lines indicate where $1728/j_\sunset(z)$ (the argument of the hypergeometric functions) is real, using the same conventions as in \cref{fig:rho24}: dotted for no cut, solid for removed cut, zigzag for remaining cut.
        Negative real values only produce a cut for $F_2(z)$ and therefore do not affect the principal branch ($\star$), which is purely $F_1(z)$.
        Positive real values only produce a cut when $|1728/j_\sunset(z)|\geq 1$, which holds within the shaded orange area.
        The monodromies described in the text are used when crossing the solid lines to keep $\rho_F(z)$ continuous: $C_0$ ($\rightarrow$) or $C_1$ ($\leftrightarrow$) is applied when going in the direction of the arrows (recall that $C_1$ is its own inverse, hence the bidirectional arrow).
        }}

\newcommand{\fignumericexamplecaption}{%
    \caption{
        Example of the process of computing $\bar E_5$ and $\bar E_6$ at $t_\ex=4.2+i\epsilon$ following \cref{eq:Ebar-detailed}, with the final summation following \cref{eq:Ebar-schematic}.
        The integration contours are shown in the $\tau$ and $\xi$ planes, which are represented as in \cref{fig:tau}.
        Filled triangles indicate a complex integration contour, and empty ones a real integral toward infinity (using $\theta$ for that along the imaginary $\tau$ axis).
        The shaded area indicates the ``too close to the cut'' zone described in the definition of $\mathcal C_\xi(\beta)$.
        For integrals using \cref{eq:hybrid-integral}, the parts from below ($<$) and above ($>$) the cutoff $\chi$ are indicated.
        All results are obtained with much higher precision than shown here.
        }}

\newcommand{\figmonodromycaption}{%
    \caption{The contours used in the monodromy construction, with cuts drawn as zigzag lines for the different functions involved.
        \textbf{Left}: $F_1(z)$, cut on $[1,\infty)$.
        \textbf{Middle}: $F_2(z)$, cut on $(-\infty,0]$.
        \textbf{Right}: $\hypergeometric abc{\frac1z}$, showing how the combination of $C_0$ and $C_1$ results in a contour that encircles the cut without crossing it.
        }}

\newcommand{\figJbubcaption}{%
    \caption[The function Jbub.]{
        The functions $\Jbub{n}(t)$, normalized so that the magnitude is roughly constant in $n$.
        Real parts are shown as solid lines, and imaginary parts as dashed lines; colors indicate different $n$ as per the plot legend.
        Below the two-pion threshold ($t=4$), the imaginary parts are identically zero and are not shown.
        Above threshold, $t$ has been given a small positive imaginary part.
        }}

\newcommand{\figmasterscaption}{%
    \caption[The six elliptic master integrals plotted as functions of $t$.]{
        Real
        (\inlinetikz{()}{
            \draw[real] (0,.5\inlinetikzheight) -- +(\legendwidth,0);})
        and imaginary
        (\inlinetikz{()}{
            \draw[imag] (0,.5\inlinetikzheight) -- +(\legendwidth,0);})
        parts of the six elliptic master integrals plotted as functions of $t$.
        The lines represent our implementations, \texttt{pySecDec} and \texttt{FeynTrop}, which (except when \texttt{FeynTrop} fails to converge) are indistinguishable on the scale of the plot.
        The isolated points
        (\inlinetikz{()}{
            \draw[real, solid] (.0  ,.5\inlinetikzheight) +(-2.5pt,0) -- +(+2.5pt,0) +(0,-2.5pt) -- +(0,+2.5pt);
            \draw[imag, solid] (.3,.5\inlinetikzheight) +(-2.5pt,0) -- +(+2.5pt,0) +(0,-2.5pt) -- +(0,+2.5pt);})
        are obtained using \texttt{AMFlow}, and the dashed lines
        (\inlinetikz{()}{
            \draw[real, series dash] (0,.5\inlinetikzheight) -- +(\legendwidth,0);})
        indicate the 8th-order series expansion around $t=0$ (we use a relatively low-order expansion here for legibility).
        The error bands are too narrow to be visible---see \cref{fig:E1,fig:reldiff}---but the absolute uncertainty of each line is shown on a logarithmic scale below each plot, using the Eisenstein variant of our implementation, as well as the truncation error of the series expansion, both for the 5th-order series shown above and for the more precise 32nd-order series used elsewhere.
        The error band of \texttt{pySecDec}, which is essentially the same for the real and imaginary parts, is shown in gray, and where applicable, that of \texttt{FeynTrop} is shown in lighter gray.
        \texttt{AMFlow} was run with a precision goal of $10^{-12}$; it provides no estimate of the actual error achieved.
        }}

\newcommand{\figEfourcaption}{%
    \caption{A dissection of $\bar E_4(4;t)$ as formed through \cref{eq:E4-finite}, showing the contributions from $E_i(2;t)$ (each multiplied by their respective polynomial in $t$) and the sum of those contributions (black), compared to the total $\bar E_4$ (gray) which also includes non-elliptic terms.
    The total contribution from the elliptic integrals to the imaginary part (black, dashed) appears to be zero, but as shown in the magnification on the right, it does in fact increase very slowly, corresponding to a gradually opening four-particle cut.
    }}

\newcommand{\figEonecaption}{%
    \caption[Symmetric log plot of the relative difference between our different implementations of $E_1(2;t)$]{
        The difference between our different implementations of $E_1(2;t)$ described in \cref{sec:E123}, as well as the 32nd-order series expansion,
        with error bands shown.
        To visualize the small differences, we use a symmetric log axis and display the relative difference between each quantity $y$ and some reference value $Y$.
        When $Y=0$ is known to be the true value
        (that is, in the region shaded as
        \inlinetikz{Xg}{
            \draw[gray, pattern=north east lines, pattern color=gray] (0,0) rectangle (\inlinetikzheight,\inlinetikzheight);}),
        we display $y$.
        Otherwise, we take $Y$ from the Eisenstein implementation and display the relative difference $\frac{y-Y}{Y}$.
        (That the Eisenstein line nevertheless deviates from zero in the non-shaded region is a rounding artefact.)
        In the bottom panel, we show the computer time needed to obtain each data point in the other two panels.\hardwarefn
    }}

\newcommand{\figreldiffcaption}{%
    \caption[Symmetric log plots of the relative difference between the three different implementations of the master integrals]{
        Symmetric log plots of the relative difference (as in \cref{fig:E1}) between our results and the benchmark methods, for all master integrals.
        The lines represent the Eisenstein variant of our method
        (\inlinetikz{X}{
            \draw[eisen] (.0cm,.5\inlinetikzheight) -- +(\legendwidth,0);
            \fill[eisen, errorband] (0,0) rectangle +(\legendwidth,\inlinetikzheight);
            }, taken as reference),
        \texttt{pySecDec}
        (\inlinetikz{X}{
            \draw[secdec] (.0cm,.5\inlinetikzheight) -- +(\legendwidth,0);
            \fill[secdec, errorband] (.0cm,0) rectangle +(\legendwidth,\inlinetikzheight);
            }),
        \texttt{FeynTrop}
        (\inlinetikz{X}{
            \draw[feyntrop] (.0cm,.5\inlinetikzheight) -- +(\legendwidth,0);
            \fill[feyntrop, errorband] (.0cm,0) rectangle +(\legendwidth,\inlinetikzheight);
            },
            only where applicable),
        \texttt{AMFlow}
        (\inlinetikz{X,}{
            \draw[gray, |-|] (.0\legendwidth, .5\inlinetikzheight) -- +(.5\legendwidth,0);
            \draw[gray,  -|] (.5\legendwidth, .5\inlinetikzheight) -- +(.5\legendwidth,0);
            }, not always visible),
        and the 32nd-order series expansion
        (\inlinetikz{X}{
            \draw[series] (.0cm,.5\inlinetikzheight) -- +(\legendwidth,0);
            \fill[series, errorband] (.0cm,0) rectangle +(\legendwidth,\inlinetikzheight);
            }).
        Recall that in the shaded region
        (\inlinetikz{Xg}{
            \draw[gray, pattern=north east lines, pattern color=gray] (0,0) rectangle (\inlinetikzheight,\inlinetikzheight);}),
        the values are shown directly rather than as relative differences.
        }}

\newcommand{\figtimecaption}{%
    \caption[Time consumption, measured in seconds per master integral evaluation, needed by the different evaluation methods for different $t$.]{
        Time needed to obtain the data points shown in \cref{fig:masters,fig:reldiff}, using the various implementations presented there.
        The series expansion is not shown, but consistently takes about 0.4 ms as in \cref{fig:E1}.
        \texttt{AMFlow} computes all $E_i(2;t)$ [or $\bar E_i(4;t)$] as one unit, and what is shown for $E_1(2;t)$ [or $\bar E_4(4;t)$] is a third of that time.\hardwarefn
        }}

\newcommand{\figPiEcaption}{%
    \caption[$\bar\Pi_E$]{
        The real (solid) and imaginary (dashed) parts of $\bar\Pi_E(t)$, and its fifth-order series expansion around $t=0$ (dotted), which has radius of convergence $4$.
        Also shown are the contributions from individual master integrals (i.e., the term in \cref{eq:PiE} consisting of that integral times a Laurent polynomial of $t$), as well as the $E_i$-independent remainder.
        The imaginary parts are only shown above their respective thresholds.
        The error bands are not visible on the plot scale. Still, the absolute uncertainties are shown on a logarithmic scale below the main panel, along with the smaller uncertainty from a 32nd-order series expansion.
        }}

\newcommand{\figPiJcaption}{%
    \caption[$\bar\Pi_J$]{
        $\bar\Pi_J(t)$, shown similarly to $\bar\Pi_E(t)$ in \cref{fig:PiE}.
        Here, the contributions are separated based on proportionality to different products of $\Jbub{n}$.
        Also shown is the purely real $\bar\Pi_\zeta(t)$ (dotted).
        The uncertainties are minuscule and are not shown.
        }}

\newlength{\tikzineqyshift}
\NewDocumentCommand{\tikzineq}{ O{} O{\tikzineqyshift} m }{%
    {%
        \tikz[%
        anchor=base,%
        baseline={([yshift=#2]current bounding box.center)},%
        #1]{#3}%
    }}

\title{The elliptic three-loop integrals of hadronic vacuum polarization in chiral perturbation theory}

\author[a]{Laurent Lellouch,}
\author[a]{Alessandro Lupo,}
\author[a]{Mattias Sjö,}
\author[b]{and Pierre Vanhove}

\affiliation[a]{Aix Marseille Univ, Université de Toulon, CNRS, CPT, Marseille, France}
\affiliation[b]{Institut de Physique Théorique, Université Paris-Saclay, CEA, CNRS,\\F-91191 Gif-sur-Yvette Cedex, France}

\abstract{  
   This work presents a detailed account of the Feynman integrals required for the three-loop hadronic vacuum polarization calculation performed in \mainref. We explain how to compute each of the three-loop integrals, and outline the mathematical framework underlying their evaluation. This culminates in a practical numerical implementation that enables fast and accurate evaluation of these integrals for arbitrary complex values of the photon virtuality.
    }

\begin{document}

\maketitle
\VerbatimFootnotes

\section{Introduction}

In the perturbative calculation of any particle physics observable, nontrivial loop integrals will eventually appear at a sufficiently high order in the perturbative series.
Here, that order is three-loop or \NNNLO\ in chiral perturbation theory~\cite{Gasser:1983yg,Gasser:1984gg}, and the observable is the hadronic vacuum polarization, a key contribution to the very low-energy QCD corrections to the magnetic moments of leptons.
We performed this calculation in a recent paper~\mainref, to which we defer the reader for a more complete picture of the physical background.
However, several novel loop integrals that arise there are sufficiently complicated that we have chosen to dedicate a paper to them.

As explained in \rmainref, vacuum polarization receives contributions from six elliptic master integrals denoted $E_i(d;t)$, evaluated in their finite two-dimensional form $E_i(2;t)$, or in the finite part $\bar E_i(4;t)$ of their four-dimensional form.
These master integrals belong to the broader class of elliptic integrals, i.e., integrals that cannot be expressed in terms of logarithms, polylogarithms, or multiple polylogarithms (i.e., iterated integrals of logarithms and rational functions).
Instead, they require elliptic functions. 

For any given loop integral, the ideal case is to have a closed-form solution, but for integrals as complicated as these, one normally has to resort to numerical methods.
Although a complete overview of the state of the art is beyond the scope of this work, we benchmark our solutions against three existing codes: 
\texttt{pySecDec}~\psdrefs, a mature general-purpose loop integrator based on sector decomposition;
\texttt{AMFlow}~\amfrefs, a more novel approach leveraging integration-by-parts relations to obtain much higher precision than \texttt{pySecDec} at the cost of much slower calculations; and 
\texttt{FeynTrop}~\ftrefs, another novel approach based on tropical decomposition, which is capable of very quickly obtaining a few digits of precision for finite integrals.
There also exist methods based on numerical solutions to the differential equations satisfied by Feynman integrals~\cite{Hidding:2020ytt, Armadillo:2022ugh, Liu:2022chg},
which require initial conditions obtained via other numerical or analytical methods, although an alternative approach that bypasses this requirement has been proposed~\cite{Ditsch:2025rdx}.
Lastly, some elliptic loop integrals are known in terms of elliptic polylogarithms, for which there exist implementations~\cite{Walden:2020odh,Duhr:2026ell} based on \texttt{GiNaC}~\cite{Bauer:2000cp}.
However, these implementations lack analytic continuation beyond the radius of convergence of the defining power series, although this has been addressed through a representation in terms of iterated integrals of Eisenstein series~\cite{Duhr:2019rrs}.

In this work, we present essentially closed-form solutions for four of the six master integrals---$E_1(2;t)$, $E_2(2;t)$, $E_3(2;t)$ and $\bar E_4(4;t)$---by extending and practically implementing the formulation of $E_1(2;t)$ in terms of elliptic trilogarithms evaluated at sixths roots of unity~\cite{Bloch:2014qca}.
While this still requires numerical calculation of this special case of elliptic polylogarithms, the speed and precision of this approach are orders of magnitude greater than those of an approach involving numerical integration.
For the remaining two master integrals---$\bar E_5(4;t)$ and $\bar E_6(4;t)$---we present an implementation in terms of contour integrals over $E_1(2;t)$.
Although it falls short of a closed-form solution, this specialized implementation offers much higher precision and stability than general-purpose codes while maintaining speed comparable to \texttt{pySecDec}.
All our results are valid for arbitrary complex values of the photon virtuality (with occasional exceptions near zero or at the two- and four-pion thresholds) and have numerical precision comfortably larger than that required by experiment.

$E_1(2;t)$ is the three-loop all-equal-masses sunset integral, and has been extensively studied.
For real virtualities below threshold, it has a Bessel integral representation~\cite{Groote:2005ay,Bailey:2008ib,Bloch:2014qca}, and for all virtualities, it has the aforementioned one in terms of elliptic trilogarithms~\cite{Bloch:2014qca}.
These are natural functions of a nome $q$ instead of the photon virtuality $t$,
\footnote{
    Note the difference between the unphysical nome, for which $q$ is the standard symbol, and the photon momentum, which is also commonly called $q$.
    To reduce the potential for confusion, here we call the photon momentum $p$ and mainly express it in terms of the photon virtuality in units of the pion mass, $t\coloneq p^2/\Mpi^2$.}
so one major component of our extension of \rcite{Bloch:2014qca} is the reliable mapping from $t$ to $q$.
Another is the straightforward implementation of $E_2(2;t)$ and $E_3(2;t)$, which relate to $E_1(2;t)$ via differentiation and therefore involve elliptic polylogarithms of lower weight.
$\bar E_4(4;t)$ is then simply a linear combination of these and non-elliptic integrals.
Lastly, we express $\bar E_5(4;t)$ and $\bar E_6(4;t)$ as integrals of $E_1(2;t)$, indicating that these would be expressible in terms of elliptic polylogarithms of higher weight, although we are unable to find this expression and therefore resort to numerical integration.

The remainder of this paper is organised as follows. 
In \cref{sec:definition} we recall the definition of the master integrals that we will evaluate numerically.
In \cref{sec:E123} we give the expressions of the masters $E_{1,2,3}$ in terms of elliptic polylogarithms as the sixth root of unity, and provide two alternative formulations of these polylogarithms in terms of Eisenstein series with a mod 6 character.
We also discuss the properties of the map between the photon virtuality $t$ and the nome $q$, needed for the numerical evaluation. 
\Cref{sec:E5E6} is devoted to the evaluation of the masters $E_5$ and $E_6$, which involves integrals over the master $E_1$.
Our results for the numerical evaluation of the master integrals and the vacuum polarization are given in \cref{sec:results}, and we conclude with \cref{sec:conclusion}.  
In \cref{app:t0-series} we give the small-$t$ expansions of the elliptic master integrals.

\section{Definitions}\label{sec:definition}

Here, we briefly restate the notation used for the integrals in \mainref, so that we may define the objects of interest.
Letting $p$ be the momentum of the external photon, with $t\coloneq p^2/\Mpi^2$ its virtuality normalized to the pion mass $\Mpi$,%
\footnote{%
    Since \rmainref\ assumes the isospin limit, there is no mass difference between charged and neutral pions.
    The common pion mass is normally taken as $\Mpi\approx135~\mathrm{MeV}$.}
we choose the basis of loop momenta
\begin{equation}\label{eq:basis}
    \{k_j\}_{j=1}^9 = \left\{
        \ell_1,\ell_2,\ell_3,\quad
        \ell_1-p,\ell_2-p,\ell_3-p,\quad
        \ell_1+\ell_3, \ell_2+\ell_3, \ell_1-\ell_2
        \right\}
\end{equation}
so that all three-loop integrals in $d$ dimensions (with $d$ an integer minus $2\epsilon$) may be written in the form
\begin{equation}
    \big[e^{\gamma_E}\Mpi^2\big]^{-3\epsilon}
    I_{\vec\nu}^{(d)}(t;\Mpi^2) = 
        \int\frac{\d^d \ell_1\; \d^d\ell_2\; \d^d\ell_3}{(2\pi)^{3d/2}} \prod_{j=1}^9 \frac{1}{\big(k_j^2 - \Mpi^2 + i\varepsilon\big)^{\nu_j}}\,,
\end{equation}
where $\vec\nu\in\Z^9$ determines which combination of propagators appears in the integral, and where the normalization factor (involving the Euler-Mascheroni constant $\gamma_E$) ensures that the integral has integer mass dimension.

Of the $11$ three-loop master integrals appearing in \mainref\ (as determined through the Laporta algorithm~\cite{Laporta:2000dsw} implemented in \texttt{LiteRed 2}~\literedrefs), four can be written as products of one-loop integrals, allowing them to be expressed entirely in terms of the bubble functions
\begin{equation}\label{eq:Jbub}
    \Jbub{n}(t) \coloneq \int_0^1 \d x\, \log^n\big[1 - x(1-x)t\big]\,.
\end{equation}
They have the closed-form expressions~\mainref[app.~C.1--2] 
\begin{subequations}
    \begin{align}
        \label{eq:Jbub-1}
        \Jbub{1}(\beta) 
            &= \beta \Big[\log(\beta_+)-\log(\beta_-)\Big]-2\,,\\[1ex]
        \label{eq:Jbub-2}
        \Jbub{2}(\beta) 
            &= 8 - 2\beta\big[f_2(\beta_+)-f_2(\beta_-)\big]\,,
            \eqnbreak\qquad
            f_2(z)\coloneq \Li_2(z)+\tfrac12\log(z)^2+2\log(z)\,,\\[1ex]
        \label{eq:Jbub-3}
        \Jbub{3}(\beta)
            &= -48 + 12\beta\big[f_3(\beta_+)-f_3(\beta_-)\big]
            +3\beta \log(\beta_+)\log(\beta_-) \big[\log(\beta_+)-\log(\beta_-)\big]\,,
            \eqnbreak\qquad
            f_3(z)\coloneq\Li_3(z)+\tfrac1{12}\log(z)^3-\tfrac{\pi^2}{12}\log(z) + f_2(z)\,,
    \end{align}
\end{subequations}
where $\beta \coloneq \sqrt{1-\frac{4}{t}}$ and $\beta_\pm\coloneq \frac{\beta\pm1}{2\beta}$.

The remaining seven master integrals are
\begin{equation}\label{eq:E}
    \begin{alignedat}{2}
        E_0(d) &\coloneq \Mpi^2 I^{(d)}_{1,1,0,0,0,0,1,1,0}(p^2,\Mpi^2) \,,\qquad&
        E_4(d;t) &\coloneq \Mpi^4 I^{(d)}_{1,1,0,1,0,0,1,1,0}(p^2,\Mpi^2) \,,\\
        E_1(d;t) &\coloneq \Mpi^2 I^{(d)}_{1,0,0,0,1,0,1,1,0}(p^2,\Mpi^2) \,,\qquad&
        E_5(d;t) &\coloneq \Mpi^6 I^{(d)}_{1,1,0,1,1,0,1,1,0}(p^2,\Mpi^2) \,,\\
        E_2(d;t) &\coloneq \Mpi^4 I^{(d)}_{2,0,0,0,1,0,1,1,0}(p^2,\Mpi^2) \,,\qquad&
        E_6(d;t) &\coloneq \Mpi^8 I^{(d)}_{2,1,0,1,1,0,1,1,0}(p^2,\Mpi^2) \,,\\
        E_3(d;t) &\coloneq \Mpi^6 I^{(d)}_{3,0,0,0,1,0,1,1,0}(p^2,\Mpi^2) \,.
    \end{alignedat}
\end{equation}
Here, additional factors of $\Mpi$ have been inserted to make the $E_i(2;t)$ dimensionless.
We state most of these integrals in terms of their finite two-dimensional forms, to which they are related through Tarasov's dimension-shifting formula~\cite{Tarasov:1996br}.
They are divergent in four dimensions, so there we use the finite part $\bar E_i(4;t)$, which we define to be dimensionless by including a factor of $\Mpi^{-6}$.

Of the masters, the three-loop sunset integral $E_1$ is the most well-studied; see, e.g.,~\rrcite{Groote:1998wy,Bailey:2008ib,Bloch:2014qca,Pogel:2022yat,Duhr:2025ouy}.
In two dimensions, it satisfies the differential equation%
\footnote{
    This is \rmainref[eqs.~(A.4--5)], but restricted to $d=2$ and divided by a common factor 4.}
\begin{equation}\label{eq:diffeq-E1}
    \bigg[
          \pi_{13} (t)\frac{\p^3}{\p t^3} 
        + \pi_{12} (t)\frac{\p^2}{\p t^2} 
        + \pi_{11} (t)\frac{\p}{\p t} 
        + \pi_{10} (t)\bigg] E_1(2;t) = 24\,,
\end{equation}
with polynomial coefficients
\begin{equation}
    \begin{alignedat}{2}
        \pi_{13}(t) &= (t-16) (t-4) t^2\,,&\qquad
        \pi_{12}(t) &= 6 t (t^2 - 15 t + 32)\,,\\
        \pi_{11}(t) &= 7t^2 - 68 t + 64\,,&\qquad
        \pi_{10}(t) &= t - 4\,,
    \end{alignedat}
\end{equation}
on which we base our calculations, as described in \cref{sec:E123}.
All other $E_i$ can be expressed in terms of $E_1$.
Trivially, $E_0(d)=E_1(d;0)$, with $E_1(2;0) = -7\zeta(3)$, where $\zeta(z)$ is the Riemann zeta function.
$E_2$ and $E_3$ are related to the $t$-derivatives of $E_1$:%
\footnote{
    This is \rmainref[eqs.~(A.6--7)], but restricted to $d=2$.}
\begin{align}
    \label{eq:E2toE1}
    E_2(2;t) &= -\frac14\bigg[t\frac{\p}{\p t} + 1\bigg] E_1(2;t)\,,\\
    \label{eq:E3toE1}
    E_3(2;t) &= \bigg[\frac{t}{2}\frac{\p^2}{\p t^2} + \frac{4+t}{8}\frac{\p}{\p t}  + \frac{1}{8} \bigg] E_1(2;t)\,.
\end{align}
Through dimension shifting, we have~\mainref[app.~C.5]
\begin{multline}\label{eq:E4-finite}
    \bar E_4(4;t)
        = \frac{t^2-20 t+160}{96} E_1(2;t) 
        - \frac{(t^2-28 t+120) (t-16) }{96} E_2(2;t)\\
        - \frac{(t-4)(t-16)^2}{48} E_3(2;t)
        + \frac{25t - 136 + 14\pi^2}{24} - \zeta(3)
        - \frac{59+3\pi^2}{8}\Jbub1(t) + \tfrac{17}{8}\Jbub{2}(t)-\tfrac{1}{4}\Jbub{3}(t)\,,
\end{multline}
so it follows directly from the evaluation of $\Jbub{1,2,3}(t)$ and $E_{1,2,3}(2;t)$.
Lastly, $\bar E_5(4;t)$ and $\bar E_6(4;t)$ can be expressed as integrals over $\Jbub{1,2,3}(t)$, and $E_1(2;t)$~\mainref[app.~C.6--7], which is the topic of \cref{sec:E5E6}.

\section{The evaluation of $E_1$, $E_2$ and $E_3$ in two dimensions}\label{sec:E123}

In \mainref[app.~D], we stated the Bessel representation of these integrals~\cite{Groote:2005ay,Bailey:2008ib}:
\begin{align}\label{eq:E1-bessel}
    E_1(2;t) &= -8\int_0^\infty x I_0(x \sqrt{t}) \big[K_0(x)\big]^4 \d x\,,\\
    E_2(2;t) &= 2\int_0^\infty x 
        \Big\{
            I_0(x\sqrt{t}) 
            + \tfrac{x\sqrt{t}}{2} I_1(x\sqrt{t})
            \Big\}
        \big[K_0(x)\big]^4 \d x\,,\label{eq:E2-bessel}\\
    E_3(2;t) &= -\frac12\int_0^\infty x 
        \Big\{
            (2+x^2) I_0(x\sqrt{t}) 
            + \tfrac{x(t+2)}{\sqrt{t}} I_1(x\sqrt{t})
            + x^2 I_2(x\sqrt{t})
            \Big\}
        \big[K_0(x)\big]^4 \d x\,,\label{eq:E3-bessel}
\end{align}
where $I_\alpha(z)$ and $K_\alpha(z)$ are the modified Bessel functions of the first and second kind, respectively.
While this representation is straightforward to write down, it is numerically costly and not valid everywhere.
Here, we will describe other, more powerful formulations due to Bloch, Kerr \& Vanhove~\cite{Bloch:2014qca} that make manifest the elliptic nature of this integral. 
Thus, this section, along with \cref{sec:t-tau-q}, is a summary of the results of~\rcite{Bloch:2014qca}, extended with the related formulae for $E_2$ and $E_3$, and with consideration given to the practical numerical evaluation of the integrals.

\subsection{Elliptic polylogarithms}\label{sec:polylog}

A natural variable for expressing the elliptic integral $E_1(2;t)$ is the \emph{nome} $q\coloneq e^{2\pi i\tau}$, where \mbox{$\Im(\tau)>0$}.
It is related to our momentum parameter $t$ through the (mirror) map
\begin{equation}\label{eq:q-to-t}
    t(q) = -\bigg[\frac{\eta(q)\eta(q^3)}{\eta(q^2)\eta(q^6)}\bigg]^6\,,
\end{equation}
where $\eta(q)$ is the Dedekind eta function defined by%
\footnote{
    In many contexts, including our numerical implementation, $\eta$ is expressed as a function of $\tau$.
    With this alternative definition, $t(\tau)=-\big[\eta(\tau)\eta(3\tau)/\eta(2\tau)\eta(6\tau)\big]^6$.}
\begin{equation}
    \eta(q) \coloneq q^{\frac{1}{24}}\prod_{n=1}^\infty\big(1-q^n\big)\,.
\end{equation}
We will return to the relationship between $t$ and $q$ (or $\tau$) in \cref{sec:t-tau-q}, but one thing to note is that since $|q|<1$, series involving $q$ tend to converge rapidly regardless of what $t$ is.

As shown in \rcite{Bloch:2014qca}, $E_1(2;t)$ is a period integral associated with a $K3$ surface defined by the singular locus of the Feynman integrand. A detailed description of this connection  
is beyond the scope of this work, but the essence is that $E_1(2;t)$ connects deeply with the geometry of that surface, endowing it with a well-behaved, relatively simple
form that is closed up to a rapidly convergent sum.%
\footnote{
    While we focus on the $d=2$ value here, there is recent work on higher-order terms in the $\epsilon$ expansion \cite{Pogel:2022yat,Duhr:2025ouy} that appear in $d=4$; other recent work on this family of integrals includes \rrcite{Duhr:2025ppd,Duhr:2025tdf,Duhr:2025kkq,Pogel:2025bca}.}

Still following \rcite{Bloch:2014qca}, $E_1(2;t)$ depends on $q$ through the \emph{period}
\begin{equation}
    \varpi_1(q) \coloneq \frac{\big[\eta(q^2) \eta(q^6)\big]^4}{\big[\eta(q) \eta(q^3)\big]^2}
\end{equation}
and the weight-$r$ \emph{elliptic polylogarithms} $\eLi_r(q,z)$ with
\begin{alignat}{2}
    \label{eq:ELi3-def}
    \eLi_3(q,z) &\coloneq
        \qquad\Li_3(z)&&+ \smash{\sum_{n\geq1}}\Big[\Li_3\big(q^n z\big)+\Li_3\big(q^n z^{-1}\big)\Big]\eqnbreak
      &&\hspace{2.7cm}+\tfrac{1}{12}(\log z)^3 - \tfrac{1}{24}\log q\,(\log z)^2 + \tfrac{1}{720}(\log q)^3\,,\\[1ex]
    \label{eq:ELi2-def}
    \eLi_2(q,z) &\coloneq \frac{\p\mathcal Li_3(q,z)}{\p\log q}
        &&= \smash{\sum_{n\geq1}} n\Big[\Li_2\big(q^n z\big)+\Li_2\big(q^n z^{-1}\big)\Big] 
        - \tfrac{1}{24} (\log z)^2 + \tfrac{1}{240} (\log q)^2 \,,\\[1ex]
    \label{eq:ELi1-def}
    \eLi_1(q,z) &\coloneq \frac{\p\mathcal Li_2(q,z)}{\p\log q}
        &&= \smash{\sum_{n\geq1}} n^2\Big[\Li_1\big(q^n z\big)+\Li_1\big(q^n z^{-1}\big)\Big]
        + \tfrac{1}{120}\log q \,,
\end{alignat}
of which we need the linear combination
\begin{equation}\label{eq:Hbball-def}
    \Hbball{r}(q) \coloneq
          24\eLi_r(q,\zeta_6)
        + 21\eLi_r(q,\zeta_6^2)
        + 8\eLi_r(q,\zeta_6^3)
        + 7\eLi_r(q,1)\,,
\end{equation}
where $\zeta_6\coloneq\exp(i\pi/3)$ is the principal sixth root of unity.
Then
\begin{align}\label{eq:E1-2d}
    E_1(2;t) &= -\varpi_1(q)\Big(40\pi^2\log q - 48\Hbball3(q)\Big)\,,
\end{align}
which is our main formula for $E_1(2;t)$.

\begin{figure}[tp]
    \centering
    \includegraphics{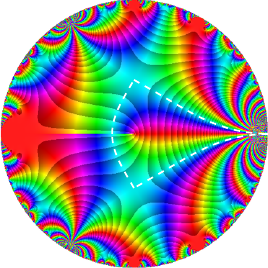}
    \includegraphics{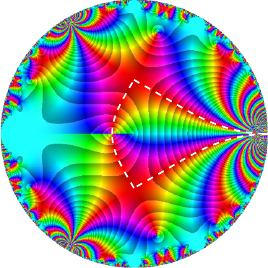}
    \includegraphics{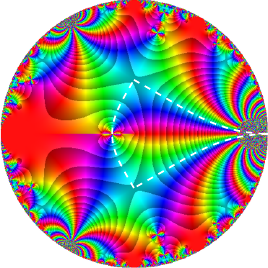}
    \figellipticcaption
    \label{fig:elliptic}
\end{figure}

The master integrals $E_2(2;t)$ and $E_3(2;t)$ are not covered in \rcite{Bloch:2014qca} but can be put in a similar form using the lower-weight elliptic polylogarithms so that, from \cref{eq:E1-2d},
\begin{align}\label{eq:E1-deriv}
   t \frac{\p}{\p t}  E_1(2;t)&= \frac
            {\Dq{\varpi_1} E_1(2;t) - \varpi_1(q)\Big(40\pi^2 - 48\Hbball2(q)\Big)}
            {\Dq t}
        \,,\\
 \Big(t   \frac{\p }{\p  t}\Big)^2E_1(2;t) &= 
        \bigg[\frac{2\Dq{\varpi_1}}{\Dq t} - \frac{\Dq[2]t}{\Dq t^2}\bigg]t\frac{\p E_1(2;t)}{\p t}
        \notag\\&\qquad
        + \frac{\big(\Dq[2]{\varpi_1}-\Dq{\varpi_1}^2\big) E_1(2;t) + 48\varpi_1(q)\Hbball1(q)}{\Dq t^2}\,,
\end{align}
where, for brevity, we have defined the differential operators
\begin{equation}\label{eq:Dq-def}
    \Dq f\coloneq \frac{\p\log f(q)}{\p\log q}\,,\qquad
    \Dq[2]f\coloneq \frac{\p^2\log f(q)}{(\p\log q)^2}\,.\qquad
\end{equation}
In the next subsection, we present an efficient way of evaluating the instances of $\Dq f$ and $\Dq[2]f$ that appear here.
Combining \cref{eq:E1-deriv} with \cref{eq:E2toE1,eq:E3toE1}, we get
\begin{align}
    \label{eq:E2-2d}
    E_2(2;t) &= 
        -\frac14\bigg[\frac{\Dq{\varpi_1}}{\Dq t} + 1\bigg] E_1(2;t) 
        + \frac{\varpi_1(q)}{4\Dq t} \Big(40\pi^2 - 48\Hbball2(q)\Big)
        \,,\\
    \label{eq:E3-2d}
    E_3(2;t) &= 
        \bigg[
            \frac{\Dq t\big(\Dq[2]{\varpi_1} + \Dq{\varpi_1}^2\big) - \Dq[2]t\Dq{\varpi_1}}{2t \Dq t^3}
            + \frac{\Dq t + \Dq{\varpi_1}}{8 \Dq t}
        \bigg] E_1(2;t)
        \notag\\&\qquad
        + \bigg[
            \frac{\Dq[2]t}{2t\Dq t^3}
            - \frac{\Dq{\varpi_1}}{t\Dq t^2}
            + \frac{1}{8\Dq t}
        \bigg]\varpi_1(q)\Big(40\pi^2 - 48\Hbball2(q)\Big)
        + \frac{24\varpi_1(q)\Hbball1(q)}{\Dq t^2}
        \,,
\end{align}
which are our main formulae for $E_2(2;t)$ and $E_3(2;t)$.
With these in hand, the integrals may be computed over the entire $q$-disk, as shown in \cref{fig:elliptic}.

\subsection{Eisenstein series with a character}\label{sec:eisenstein}
While the sums in \cref{eq:ELi3-def,eq:ELi2-def,eq:ELi1-def} converge well, the functions of $q$ used here possess a form that is even better suited for numerical evaluation.
This involves \emph{Eisenstein series}, which we express as%
\footnote{
    This notation is \emph{ad hoc}, with the conventional Eisenstein series being (for $k\geq2$)
    \begin{equation*}
        E_{2k}(\tau)\coloneq \frac{1}{2\zeta(2k)}\sum_{(m,n)\neq(0,0)}\frac{1}{(m+n\tau)^{2k}} = 1 + \frac{2\E(1-2k,1;q)}{\zeta(1 - 2k)}\,.
    \end{equation*}}
\begin{equation}\label{eq:eisenstein-def}
    \E(a,b;q) \coloneq \sum_{n\geq 1} \frac{1}{n^a}\frac{q^n}{(1 - q^n)^b}\,.
\end{equation}
Such series converge rapidly.
To compactly represent linear combinations of the form $\sum_{k\in K}c_k\E(a,b;q^k)$ where $K $ and $\{c_k\}_{k \in K}$ are sets of integers, we use
\begin{equation}\label{eq:eisenchar-def}
    \E(a,b,\psi;q) \coloneq \sum_{n\geq 1} \frac{\psi(n)}{n^a}\frac{q^n}{(1 - q^n)^b}\,,
    \qquad
    \psi(n) \coloneq \sum_{k\in K} 
        \begin{cases}
            c_k k^a & \text{if $k$ divides $n$,}\\ 
            0       & \text{otherwise,} 
        \end{cases} 
\end{equation}
where the \emph{character} $\psi(n)$ is periodic with a period equal to the least common multiple of $K$.
All cases appearing here have period 6: $\psi(n+6)=\psi(n)$.

As proven in \rcite{Bloch:2014qca},%
\footnote{
    While deriving either form of $\Hbball3(q)$ from the Feynman integral involves arguments from Hodge theory, which is beyond the scope of this paper,
    it is not too difficult to show that the two are equivalent:
    starting from \cref{eq:Hbball3-eisen}, split $\E(3,1,\psi_\bball;q)$ into a linear combination of $\E(3,1;q^k)$, Taylor expand $1/(1-q^n)$, and perform the sum over $n$  using $\Li_r(z)=\sum_{n\geq 1} z^n/n^r$ to arrive at \cref{eq:Hbball-def}.}
\begin{equation}\label{eq:Hbball3-eisen}
    \Hbball3(q) = \frac{(\log q)^3}{12} + \frac{5\pi^2\log q}{6} - \frac{\zeta(3)}{3} + \E(3,1,\psi_\bball; q)\,, 
\end{equation}
where 
$\{\psi_\bball(n)\}_{n=1}^6 = \{1,-15,-8,-15,1,120\}$.
Other examples, easily obtainable from the definitions, are
\begin{equation}\label{eq:eisen-examples}
    \begin{alignedat}{2}
        \log\eta(q) &= \frac{\log q}{24} - \E(1,1; q)\,,&&\\
        \log[-t(q)] &= - \log q - 6\E(1,1,\psi_t; q)\,,\qquad&
            \{\psi_t(n)\}_{n=1}^6 &= \{1,-1,4,-1,1,-4\}\,,\\
        \log\varpi_1(q) &= \log q + 2\E(1,1,\psi_\varpi; q)\,,\qquad&
            \{\psi_\varpi(n)\}_{n=1}^6 &= \{1,-3,4,-3,1,-12\}\,.
    \end{alignedat}
\end{equation}
Derivatives are straightforwardly obtained through
\begin{equation}
    \frac{\p\E(a,b,\psi;q)}{\p\log q} = b\E(a-1,b+1,\psi; q) + (1-b)\E(a-1,b,\psi; q)\,,
\end{equation}
which readily provides 
\begin{equation}\label{eq:Dqt-eisen}
    \Dq{t}=-1-6 \E(0,2,\psi_t;q)\,,\qquad
    \Dq{\varpi_1}=1+2\E(0,2,\psi_\varpi;q)\,,
\end{equation}
etc., as needed for efficient evaluation of $E_2(2;t)$ and $E_3(2;t)$ via \cref{eq:E2-2d,eq:E3-2d}.
Furthermore, from \cref{eq:Hbball3-eisen},
\begin{align}
    \label{eq:Hbball2-eisen}
    \Hbball2(q) = \frac{\p\Hbball3(q)}{\p\log q}
        &= \frac{(\log q)^2}{4} + \frac{5\pi^2}{6} + \E(2,2,\psi_\bball;q)\,,\\
    \label{eq:Hbball1-eisen}
    \Hbball1(q) = \frac{\p\Hbball2(q)}{\p\log q}
        &= \frac{\log q}{2} + \E(1,3,\psi_\bball;q) - \E(1,2,\psi_\bball;q)\,.
\end{align}

Closed-form expressions sometimes exist.
For instance, using
\cite[sequence~A125510]{oeis} we can express  $\Dq{t}$ as a sum of $\eta$-ratios: 
\begin{equation}\label{eq:Dqt-eta}
    \Dq{t} = 
        -\frac{
            \big[\eta(q)\eta(q^2)\big]^4 + 9\big[\eta(q^3)\eta(q^6)\big]^4
        }{
            \eta(q)\eta(q^2)\eta(q^3)\eta(q^6)
        }\,.
\end{equation}

\subsection{Poisson resummation}\label{sec:poisson}

The formulae provided above converge poorly near $t=0$, corresponding to $q$ near $1$.
While the series expansions of the integrals (\cref{app:t0-series}) are accurate in this limit, it is also possible to improve the behavior of the elliptic formulation through Poisson resummation.
The steps follow \rcite[\S2.4]{Bloch:2014qca}.

Starting from the Eisenstein series formulation \cite[theorem 2.2]{Bloch:2014qca},
\begin{equation}\label{eq:E1-eisen}
    E_1(2;t) = -\varpi_1(\tau) \bigg[16\zeta(3) - 48\sum_{n\geq 1}\frac{\psi_\bball(n)}{n^3}\frac{q^n}{1-q^n} + 32i\pi^3\tau^3\bigg]
\end{equation}
[compare \cref{eq:E1-2d,eq:Hbball3-eisen}],
we apply \rcite[lemma 2.4.1]{Bloch:2014qca} followed by some algebraic manipulation:
\begin{align}
    16\zeta(3)-48\sum_{n\geq1} \frac{\psi_\bball(n)}{n^3} \frac{q^n}{1-q^n}
        &= \frac{\tau}{2\pi i} \sum_{m\in\Z}\sum_{n\geq1} \frac{\psi_\bball(n)}{n^2} \frac{1}{m^2-(n\tau)^2}\notag\\
        &= \frac{\tau}{2\pi i} \sum_{m\in\Z}\sum_{n\geq1} \frac{\psi_\bball(n)}{m^2} \frac{1}{m^2-(n\tau)^2}\notag\\
        &= \frac{\tau}{2\pi i} \sum_{m\geq1}\sum_{n\in\Z} \frac{\psi_\bball(n)}{m^2} \frac{1}{m^2-(n\tau)^2} - 32i\pi^3\tau^3\,,
\end{align}
where in the second step we used the partial fraction identity
\begin{equation}
    \frac{1}{x^2(y^2-x^2)} = \frac{1}{y^2(y^2-x^2)} + \frac{1}{x^2y^2}
\end{equation}
along with $\sum_{n\geq1} \frac{\psi_\bball(n)}{n^2}=0$ to remove the second term.
Substituting back into \cref{eq:E1-eisen} and exploiting the periodicity of $\psi_\bball(r)$, we may write
\begin{equation}\label{eq:E1-lattice}
    E_1(2;t) = -\varpi_1(\tau)\, \frac{\tau^3}{2\pi i}\sum_{r=1}^6\sum_{m\geq1}\sum_{n\in\Z} \frac{\psi_\bball(r)}{m^2} \frac{1}{m^2-\big[(r+6n)\tau\big]^2}
\end{equation}
and perform the Poisson resummation with respect to $n$:
\begin{equation}
    \sum_{n\in\Z} \frac{1}{m^2-\big[(r+6n)\tau\big]^2}
        = \sum_{\hat n\in Z} \int_{-\infty}^\infty \frac{e^{-i\hat n x\pi}\,\d x}{m^2-\big[(r+6n)\tau\big]^2}
        = \frac{i\pi}{6m\tau}\sum_{\hat n\in\Z} e^{-i\pi \frac{m|\hat n|}{3\tau} + i\pi\frac{\hat n r}{3}} \,.
\end{equation}
The sums over $m$ now have closed-form solutions, so substituting back, we get
\begin{equation}\label{eq:E1-poisson}
    E_1(2;t)=-\varpi_1(\tau)\, \bigg[
        -8\tau^2 \sum_{\hat n\geq 1}\Li_3\Big(e^{-i\pi\frac{\hat n}{3\tau}}\Big)\sum_{r=1}^6 \psi_\bball(r) \cos\Big(\frac{\pi \hat n r}{3}\Big)
        -336\zeta(3)\tau^2\bigg]\,.
\end{equation}
Compared to the other formulations, this is numerically better behaved as $\tau\to 0$.
Although $e^{-i\pi\frac{\hat n}{3\tau}}$ has an essential singularity at $\tau=0$, its pathological behavior does not materialize when $\tau$ approaches zero along the imaginary axis, which corresponds to $t$ approaching zero from below (see \cref{fig:tau}). 
When $t$ approaches zero from above, $\tau$ approaches zero along a circular arc tangent to the imaginary axis, which puts it close enough to the axis that the behavior remains non-pathological.

Poisson-resummed formulae for $E_2(2;t)$ and $E_3(2;t)$ can be read off from \cref{eq:E2-2d,eq:E3-2d} by noting that the quantity in brackets in \cref{eq:E1-poisson} is \smash{$\big(40\pi^2\log q - 48\Hbball3(q)\big)$} [compare \cref{eq:E1-eisen}], so its first $\tau$-derivative is \smash{$2\pi i\big(40\pi^2 - 48\Hbball2(q)\big)$}, and the second, \smash{$192\pi^2\Hbball1(q)$}. 
  
\subsection{Obtaining $q$ and $\tau$ from $t$}\label{sec:t-tau-q}

\begin{figure}[p]
    \begin{subfigure}[t]{.6\columnwidth}
        \hspace{-.7cm}
        \includegraphics{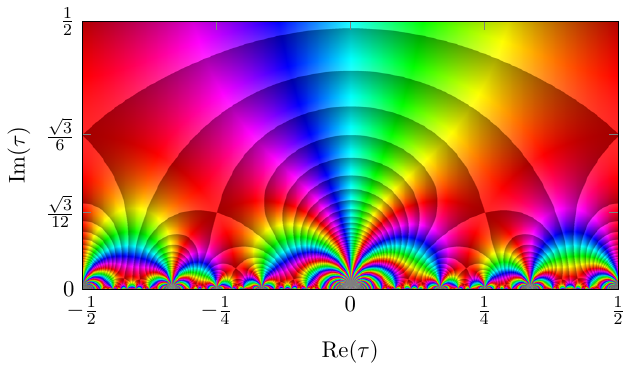}
    \end{subfigure}
    \hspace{.7cm}
    \begin{subfigure}[t]{.3\columnwidth}
        \includegraphics{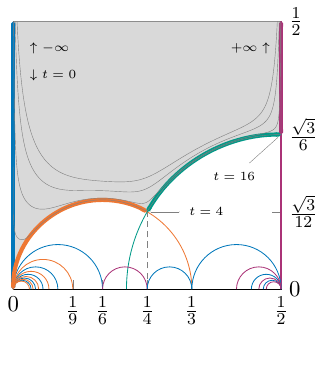}
    \end{subfigure}
    \figtaucaption
    \label{fig:tau}
\end{figure}

\begin{figure}[p]
    \centering
    \begin{subfigure}[t]{.45\columnwidth}
        \includegraphics{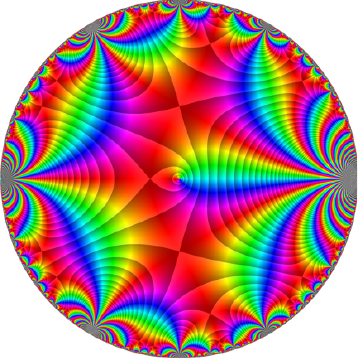}
    \end{subfigure}
    \hspace{1cm}
    \begin{subfigure}[t]{.45\columnwidth}
        \includegraphics{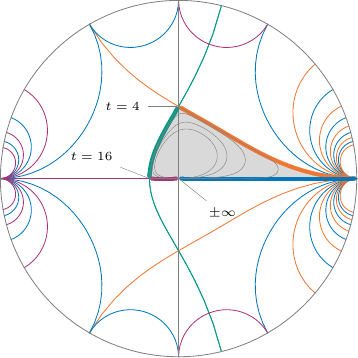}
    \end{subfigure}
    \figqcaption
    \label{fig:q}
\end{figure}

While the expressions for $E_i(2;t)$ presented above are analytically and numerically well-behaved,
they are given in terms of $q$, whose complicated relation \cref{eq:q-to-t} with the kinematic variable $t$ means it is not \emph{a priori} clear how to evaluate them for a given value of $t$.
In this section, aided by \cref{fig:tau,fig:q}, we aim to clarify that relationship, allowing for the direct evaluation of $E_1(2;t)$ anywhere in the complex $t$-plane.

The nome $q$ lives in the unit disk $|q|<1$, and its parameter $\tau$ lives in the upper half-plane $\Im(\tau)>0$.
The mapping 
$\tau\mapsto q = e^{2\pi i\tau}$ 
maps the real axis to the boundary of the disk, and $+i\infty$ to its center.
It has period $1$ in the real direction, so $\tau$ can be restricted to the strip $-\frac12<\Re(\tau)\leq\frac12$ without loss of generality.
With this restriction, the mapping is one-to-one.
The presence of $\log q = 2\pi i\tau$ in $E_1(2;t)$ makes it sensitive to $\tau \mapsto \tau+n$.
This is as expected: because $E_1(2;t)$ is a Feynman integral, it should not be a modular form (which would be invariant under $\tau\mapsto\tau+n$) but a multivalued function whose monodromy (discontinuity across its branch cut) corresponds precisely to $\tau\mapsto\tau\pm1$.

\afterpage{\clearpage} 

The mapping 
$q\mapsto t$ 
[i.e., \cref{eq:q-to-t}]
is ill-behaved near any point on the rim of the disk where every possible value of $t$ (except $0$, which never occurs) appears infinitely many times.
On the other hand, $t=0$ is obtained without problem in, e.g., the limit $q\to1$ along the real axis ($\tau\to0$ along the imaginary axis).
Every nonzero value of $t$ appears infinitely often in the image of the upper half-plane under 
$q\mapsto t$,
as illustrated in the fractal-like structure of \cref{fig:tau,fig:q}.
In these figures, we have indicated a region 
that contains each $t$ with $\Im(t)>0$ exactly once,
and whose boundary is a line (bold) corresponding to the real $t$-axis.

The inverse mapping 
$t\mapsto q$ 
(or rather $t\mapsto\tau$) 
is described in \rcite[\S3.2]{Bloch:2014qca}.
However, it involves several multivalued functions (denoted $\rho$ below) which, if restricted to their respective principal branches, fail to produce the correct result: it is entirely wrong for $-0.12\lesssim t \lesssim 0.2$ and correct only up to a phase (a piecewise constant power of $i$) elsewhere.
Here, we correct the branch choices so that the mapping $t\mapsto\tau$ is analytic everywhere except for a branch cut on the real line.%
\footnote{
    This procedure of correcting $f(g(z))$, where $f$ is multivalued and $g$ non-monotonous, by supplanting $f$ with information about $z$, can be thought of as analogous to how $\mathrm{atan2}(y,x)$ (as implemented by most programming languages) is the correct way of computing $\arg(x+iy)$, whereas $\tan^{-1}(y/x)$ may be wrong by a multiple of $\pi$.}

\begin{figure}[tp]
    \begin{subfigure}[t]{.45\columnwidth}
        \hspace{-.7cm}
        \includegraphics{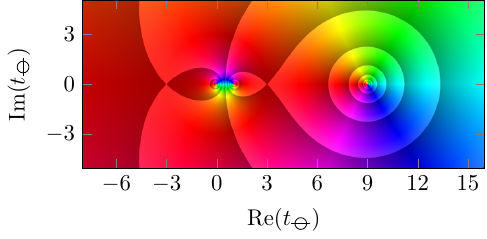}
    \end{subfigure}
    \hspace{.7cm}
    \begin{subfigure}[t]{.45\columnwidth}
        \includegraphics{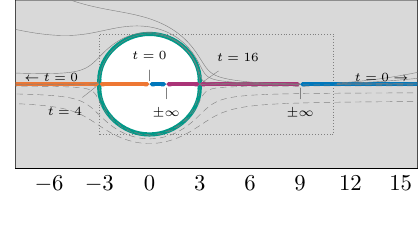}
    \end{subfigure}
    \figtsuncaption
    \label{fig:tsun}
\end{figure}

The mapping is, as shown in \rcite{Bloch:2013tra}, given by a ratio of periods,%
\footnote{
    The factor of $i$ is strictly speaking not part of the relation, and can be absorbed by appropriately redefining the $\rho$ functions below.
    However, placing it here allows us to make their definition the same for $\Im(t)>0$ and $\Im(t)<0$, simplifying the expressions.}
\begin{equation}\label{eq:t-to-tau}
    \tau(t)
    = i\frac{\varpi_c(t_\sunset)}{\varpi_r(t_\sunset)}\,,\qquad
    t = -\frac{64 t_\sunset}{(t_\sunset - 9)(t_\sunset - 1)}\,,
\end{equation}
where $t_\sunset$ is the kinematic parameter of the underlying two-loop sunset integral and is visualized in \cref{fig:tsun}.
The periods are
\begin{equation}\label{eq:periods}
    \varpi_c(z) \coloneq \frac{2\pi \rho_F(z)}{\sqrt[4]{z-3} \rho_4(z)}\,,\qquad
    \varpi_r(z) \coloneq \frac{12\sqrt{3}}{z-9} \,\varpi_c\bigg(\frac{9z-9}{z-9}\bigg)\,,
\end{equation}
where $\rho_4(z)$ and $\rho_F(z)$ are nominally
\begin{equation}\label{eq:rho4F}
    \rho_4(z) = \sqrt[4]{z^3 - 9z^2 + 3z - 3}\,,\qquad 
    \rho_F(z) = F_1\big(1728/j_\sunset(z)\big)
\end{equation}
where
\begin{equation}\label{eq:F1}
    F_1(z)\coloneq \hypergeometric[\bigg]{\sfrac1{12}}{\sfrac5{12}}{1}{\,z}\,,\qquad
    j_\sunset(z) \coloneq \frac{(z-3)^3(z^3-9z^2+3z-3)^3}{(z-9)(z-1)^3 z^2}\,,
\end{equation}
and $\hypergeometric abcz$ is the Gauss hypergeometric function.
However, an appropriate, nontrivial choice of branch must be made for these multivalued functions to obtain the correct mapping $t \mapsto\tau$. 
We will therefore replace them by canonical versions that are analytic and single-valued when restricted to a subset $\Omega$ of the complex plane, such that there is a one-to-one mapping between this $\Omega$ and the entire $t$-plane.

\begin{figure}[tp!]
    \centering
    \hspace{-.7cm}
    \includegraphics{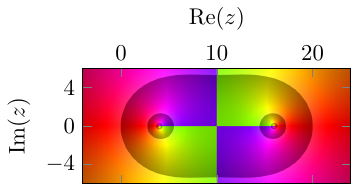}
    \hspace{-.5cm}
    \includegraphics{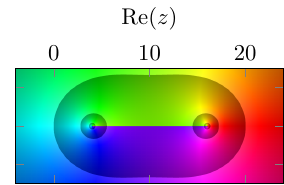}
    \hspace{-.5cm}
    \includegraphics{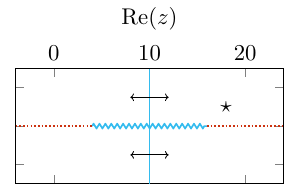}

    \hspace{-.7cm}
    \includegraphics{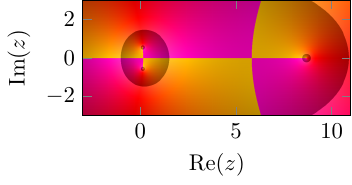}
    \hspace{-.5cm}
    \includegraphics{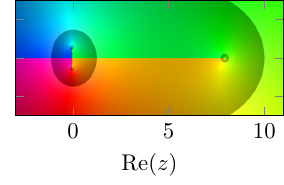}
    \hspace{-.5cm}
    \includegraphics{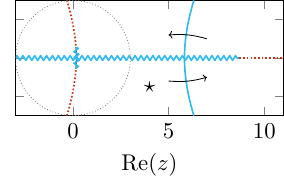}
    \figrhotwofourcaption
    \label{fig:rho24}
\end{figure}

Looking at \cref{eq:t-to-tau,fig:tsun}, we see two candidates for such a subset: $t_\sunset\mapsto t$ is a double cover, mapping each of
\begin{equation}
    \begin{aligned}
        \Omega_< &\coloneq \big\{t_\sunset\in\mathbb C\;\big|\;\abs(t_\sunset)< 3\big\} \setminus [1,3)\setminus (3,0]\,,\\
        \Omega_> &\coloneq \big\{t_\sunset\in\mathbb C\;\big|\;\abs(t_\sunset)> 3\big\} \setminus (-\infty,9]
    \end{aligned}
\end{equation}
to a separate copy of $\mathbb C\setminus[0,\infty)$.%
\footnote{%
    Note that we exclude $[0,\infty)$ from the $t$-plane, even though the branch cuts of the master integrals only start at $t=4$ or $t=16$.
    This is partly because we expect a discontinuous $t\mapsto\tau$ map at $t\in[0,16]$---compare \cref{fig:tau}, where these parts of the boundary of the $\Im(t)>0$ region are not adjacent to their $\Im(t)<0$ counterparts---and partly because it is not possible to make the mapping analytic on more inclusive sets like $\mathbb C\setminus[4,\infty)$ or $\mathbb C\setminus[16,\infty)$.
    As can be seen in \cref{fig:rho24}, $\rho_4$ will inevitably have a branch cut on the parts of the real line corresponding to either $t<0$ or $0<t<4$ (or somewhere worse), and as can be seen in \cref{fig:tsun}, the step from $t+i\epsilon$ to $t-i\epsilon$ for $t\in(4,16)$ is not local in $\Omega_>$.}
The same is true in the context of $\varpi_r$, since $z\mapsto\frac{9z-9}{z-9}$ maps each of $\Omega_\gtrless$ to itself.
However, $\rho_4$ and $\rho_F$ possess branch points at the complex roots $z\approx 0.15\pm0.57i$ of $z^3-9z^2+3z-3$, which are in $\Omega_<$ and therefore make it impossible to construct single-valued versions there.
The real root $z\approx 8.69$ is also a branch point, but is not contained in $\Omega_>$ and therefore poses no problem.
Thus, we will focus our construction on $\Omega_>$.%
\footnote{
    $\Omega_>$ has the downside of not being bounded, with $t=0$ corresponding to $t_\sunset=\infty$, but this is a small price to pay for single-valuedness.}

Let us then invert
$t_\sunset\mapsto t$ 
such that we map onto $\Omega_>$.
From \cref{eq:t-to-tau},
\begin{equation}\label{eq:t-to-tsun}
    t_\sunset = \frac{5t - 32 + 4\rho_2(t)}{t}\,,
\end{equation}
where $\rho_2$ is nominally equal to $\sqrt{(z-4)(z-16)}$, but to obtain the correct mapping, we must make the correct choice of branch (i.e., sign in front of the square root).
Because $\sqrt{z}$ has a branch cut along the negative real axis,  $\rho_2(z)$
has a branch cut at $z\in[4,16]$, which we retain, and one along $\Re(z)=10$, which we remove through (see \cref{fig:rho24}):
\begin{equation}\label{eq:rho2}
    \rho_2(z) \coloneq \sqrt{(z-4)(z-16)} \times 
    \begin{cases}
        -1  &   \text{if $\Re(z)<10$},\\
        +1  &   \text{if $\Re(z)>10$}.\\
    \end{cases}
\end{equation}
This makes $\rho_2(t)$ analytic on $\mathbb C\setminus[4,\infty)$ and makes \cref{eq:t-to-tsun} map $t$ onto $\Omega_>$; the opposite sign choice would map it onto the undesirable $\Omega_<$.

Now we can turn to $\rho_4(z)$.
Its branch cuts happen when \mbox{$\Im(z^3 - 9t^2 + 3t - 3)$} changes sign, which happens on the real axis and on the hyperbola \mbox{$\{x+iy\,|\, y^2=x^2-18x+3\}$} (see \cref{fig:rho24}).
Of course, there is only a branch cut on the segments of these curves where \mbox{$\Re(z^3 - 9t^2 + 3t - 3)<0$}, but by assigning different phases to each of the six regions into which these curves divide the complex planes, we can change along which segments the discontinuities appear in practice.
If done appropriately, this gives a result that is analytic and single-valued on $\Omega_>$.
Of course, $\rho_4(z)$ is not uniquely determined by this condition, but the global phase can rather easily be fixed by seeing which of the four possibilities gives the correct $t\mapsto\tau$.
The appropriate choice is
\begin{equation}\label{eq:rho4}
    \rho_4(z) \coloneq \sqrt[4]{z^3 - 9z^2 + 3z - 3} \times
    \begin{cases}
        +i  &   \text{if $\Im(z^3 - 9z^2 + 3z - 3)<0$ and $\Re(z)>3$,}\\
        -1  &   \text{if $\Im(z)>0$,}\\
        +1  &   \text{if $\Im(z)<0$,}
    \end{cases}
\end{equation}
where the complicated first condition means ``to the right of the hyperbola''.
This gives $\rho_4(z)$ branch cuts along the real axis below the real root of $z^3 - 9z^2 + 3z - 3$, and between the complex roots along the hyperbola (see \cref{fig:rho24}).

Lastly, we have $\rho_F(z)$.
Looking at \cref{eq:F1}, $F_1(z)$ has a branch cut along $(1,\infty)$, producing numerous discontinuities, and to treat it like we just treated $\rho_2$ and $\rho_4$, we must first determine its monodromy (behavior when moving between branches), which is less simple than just picking up phases.
Helpful formulae for this task are found in 
\rrcite{Bateman1953:htf,Lebedev:1972sfa}.

\begin{figure}[tp]
    \centering
    \includegraphics{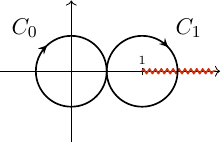}
    \quad
    \includegraphics{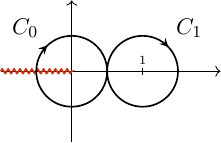}
    \quad
    \includegraphics{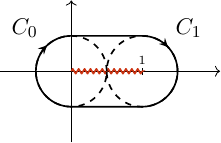}
    \figmonodromycaption
    \label{fig:monodromy}
\end{figure}

$F_1(z)$ is a solution of the hypergeometric differential equation \cite[\S9.5(9.5.4)]{Lebedev:1972sfa} with $(a,b,c)=(\sfrac1{12},\sfrac5{12},1)$,
\begin{equation}\label{eq:hyp-ode}
    0 = 144z (z-1)f''(z) +  \left(216 z -144\right)f'(z) + 5 f(z) \,,
\end{equation}
which has as its other independent solution \cite[\S9.5(9.5.6)]{Lebedev:1972sfa}
\begin{equation}
    F_2(z)\coloneq\hypergeometric[\bigg]{\sfrac1{12}}{\sfrac5{12}}{\sfrac12}{1-z}\,.
\end{equation}
Since $\hypergeometric abcz$ has a branch cut along $z\in[1,\infty)$, $F_1(z)$ has the same cut whereas $F_2(z)$ has its cut along $z\in(-\infty,0]$.
Letting $C_x$ be the clockwise circular contour of radius $1/2$ centered on $z=x$, we will therefore consider the behavior of $F_1(z)$ and $F_2(z)$ when analytically continuing around $C_0$ and $C_1$, i.e., their monodromy around the points $0$ and $1$, respectively.
This is illustrated in \cref{fig:monodromy}.

The monodromies are a linear transformation in the space
spanned by   $F_1(z)$ and  $F_2(z)$:
\begin{align}\label{e:Monodromies}
  F_1(z) &\transf{C_0} \alpha F_1(z) +\beta F_2(z)\,,\qquad
    F_2(z)\transf{C_0} \gamma F_1(z)+\delta F_2(z)\,,\cr
     F_1(z)& \transf{C_1} \tilde \alpha F_1(z) +\tilde \beta F_2(z)\,,\qquad
    F_2(z)\transf{C_1} \tilde \gamma F_1(z)+\tilde \delta F_2(z)\,,
\end{align}
where the constants $\alpha, \beta,\gamma,\delta$  are determined by the local behavior of
the functions $F_1(z)$ and $F_2(z)$ near $z=0$, and the constants
$\tilde \alpha,\tilde \beta,\tilde \gamma,\tilde \delta$ by the local behavior near $z=1$.

Following the Frobenius method (see, e.g., \rcite[XVI(\S16.1)]{Ince:1956ode}), \cref{eq:hyp-ode} has one analytic solution in the vicinity of $0$ as well as one containing $\log(z)$.
Since $F_1(z)$ has no branch cut near $0$, the logarithmic component must be entirely contained within $F_2(z)$, and working out the solution by series expansion (as is done explicitly in \rcite[\S9.7(9.7.5)]{Lebedev:1972sfa}), we find
\begin{equation}\label{eq:F2-near0}
    F_2(z) \overset{z\sim 0}{=} -\frac{ \log(z) }{4\pi}\frac{F_1(z)}{F_1(1) } + \text{(analytic terms)}\,.
\end{equation}
Continuing along $C_0$ results in $\log(z)\to\log(z) + 2\pi i$, so the $C_0$ monodromy is the linear transformation%
\footnote{ 
    This can also be seen from the identity \cite[\S2.10(1)]{Bateman1953:htf}
    \begin{equation*}
        F_1(z)
            = \frac{\Gamma(\sfrac12)}{\Gamma(\sfrac{11}{12})\Gamma(\sfrac{7}{12})} F_2(z)
            + \frac{\Gamma(\sfrac{{-1}}2)}{\Gamma(\sfrac{1}{12})\Gamma(\sfrac{5}{12})} F_3(z)\,,
    \end{equation*}
    where $F_3(z)\coloneq \sqrt{1-z}~\hypergeometric[\big]{\sfrac{11}{12}}{\sfrac{7}{12}}{\sfrac32}{1-z}$
    and, incidentally, the coefficient in front of $F_2(z)$ is equal to $F_1(1)$ \cite[\S2.8(46)]{Bateman1953:htf}.
    The right-hand side is unchanged by $C_1$ except that the square root picks up a minus sign, so eliminating $F_3(z)$, we find \cref{eq:monodromy-C1}.}
\begin{equation}\label{eq:monodromy-C0}
    F_1(z) \transf{C_0} F_1(z)\,,\qquad
    F_2(z) \transf{C_0} F_2(z) - \frac{iF_1(z)}{2F_1(1)}\,.
\end{equation}
Note that repeating $C_0$ produces an infinite tower of branches, although none of these is directly accessible from the principal sheet of $\rho_F(z)$, which is purely $F_1(z)$ and therefore unaffected by $C_0$.

\begin{figure}[p!]
    \centering
    \hspace{-.5cm}
    \includegraphics{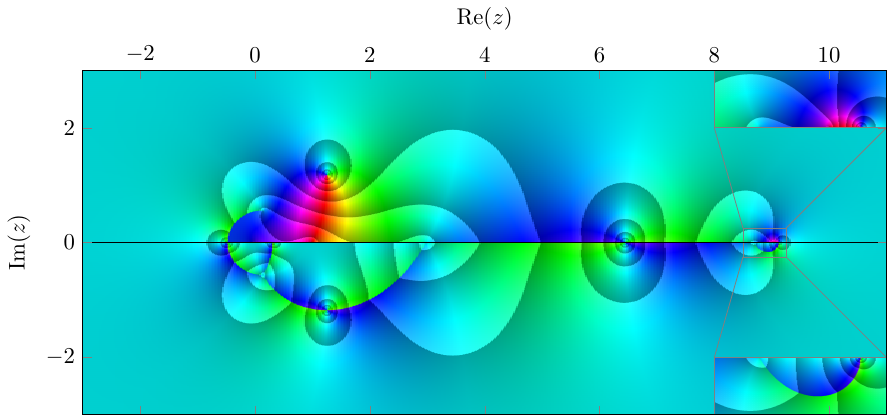}

    \hspace{-.5cm}
    \includegraphics{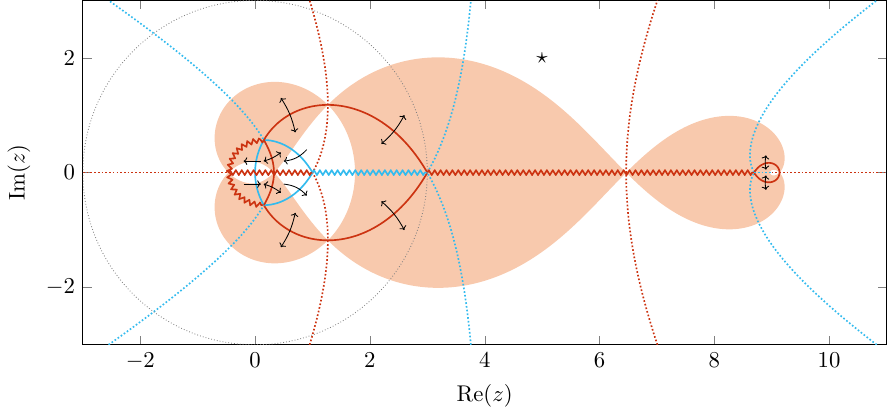}
    \figrhoFcaption
    \label{fig:rhoF}
\end{figure}
\afterpage{\clearpage} 

Again, following the Frobenius method,
\cref{eq:hyp-ode} has one analytic solution in the vicinity of $1$ as well as one containing $\sqrt{1-z}$.
Since $F_2(z)$ has no branch cut near $1$, the square root component must be contained within $F_1(z)$, and through series expansion (done explicitly in \rcite[\S9.7(9.7.1)]{Lebedev:1972sfa}) we find
\begin{equation}\label{eq:F1-near1}
    F_1(z) \overset{z\sim 1}{=} F_1(1)F_2(z) + \sqrt{1-z}\times\text{(analytic terms)}\,,
\end{equation}
which also follows from \rcite[\S9.7(9.7.1)]{Lebedev:1972sfa}.
The effect of $C_1$ is to change the sign of the square root, so eliminating the analytic terms, we find
\begin{equation}\label{eq:monodromy-C1}
    F_1(z) \transf{C_1} 2F_1(1)F_2(z) - F_1(z)\,,\qquad
    F_2(z)\transf{C_1} F_2(z)\,.
\end{equation}
Note that, unlike $C_0$, $C_1$, only alternates between two branches: $C_1^2 = 1$.

With the monodromy in hand, we can construct $\rho_F(z)$ similarly to $\rho_2(z)$ and $\rho_4(z)$, as illustrated in \cref{fig:rhoF}. 
An explicit result in the style of \cref{eq:rho2,eq:rho4} would be long and not very illuminating, so we defer it to the attached code~\cite{hvpnumerics}.
It has a branch cut along $(-\infty,9)$, and like $\rho_4(z)$, it also has one between the complex roots of $z^3 - 9z^2 + 3z - 3$, this time following the leftmost bend of the cubic curve \mbox{$\{x+iy\,|\,y^2(4-x) = x^3-6x^2+9x\}$}.%
\footnote{
    If one crosses this cut with a $C_0$ monodromy, one recovers the principal branch times $-i$.
    This is because $(C_0C_1)^3=-i$, which can be understood from the identities \cite[\S2.10(2--3)]{Bateman1953:htf} 
    that relate both $F_1$ and $F_2$ to various $\hypergeometric abc{\frac1z}$ times $12$th roots.
    Those hypergeometric functions only have a branch cut along $(0,1)$, so the combined $C_0C_1$ contour (see \cref{fig:monodromy}) is equivalent to a loop that encircles the entire cut without crossing it, only picking up phases from the $12$th roots.
    After three laps around the combined contour, these yield a global factor of $-i$.
    This can be used to derive \cref{eq:monodromy-C1} without the Frobenius method, and
    can also be shown based on the integral representation of the hypergeometric function \cite[\S4.2]{Vanhove:2018elu}.}

\subsection{The relation between variables}
We have a large number of different variables ultimately describing the same thing as the photon virtuality $t$: the velocity $\beta$, in terms of which we describe $\Jbub{n}$ as well as the differential equations solved in \cref{sec:E5E6}; and the nome $q$ and its parameter $\tau$, in terms of which we describe $E_1$, $E_2$ and $E_3$ in this section.
Going forward, we will use all of these simultaneously, so for clarity, we will 
use $\map{x}{y}$ as shorthand for the function $f:x\mapsto y$
that maps variable $x$ to variable $y$.
The concrete implementations of this function are summarized in \cref{tab:map}.

For future convenience, we will add yet another parameter, $\theta$, which parametrizes the line in $\tau$-space that corresponds to real $t$:
\begin{equation}\label{eq:theta}
    \map{\theta}{\tau}(\theta) = 
        \begin{cases}
            -\frac{i\theta}{16} &   \text{if }\theta < 0\,,\\
            \tfrac16 + \tfrac16e^{i\pi\frac{6-\theta}{6}}  &   \text{if }0 \leq \theta < 4\,,\\
            \tfrac12 + \tfrac{\sqrt3}{6}e^{i\pi \frac{34-\theta}{36}}   &   \text{if }4 \leq \theta < 16\,,\\
            \tfrac12 + \tfrac{i\theta\sqrt3}{96}                        &   \text{if }16 \leq \theta\,.
        \end{cases}
\end{equation}
Composed with $\map{\tau}{t}$, this makes $\map{\theta}{t}$ a monotonous, once continuously differentiable function with the property that $\map{\theta}{t}(\theta)=\theta$ for $\theta\in\{0,4,16\}$.
This parametrization of the real $t$-axis makes up for the fact that our implementation of $\map{t}{\tau}$ requires $\Im(t)>0$.

\begin{table}[t]
    \hspace{-.7cm}
    \begin{tabular}{c|ll|ccccc}
        \toprule
        $x$         &   Domain              &   Describes                       &  $\map{t}{x}(t)$  &   $\map{\beta}{x}(\beta)$ &   $\map{q}{x}(q)$ &   $\map{\tau}{x}(\tau)$   &   $\map{\theta}{x}(\theta)$   \\
        \midrule
        $t$        &   $\mathbb C$         &   $\bar\Pi_T$                      &   $t$             &   $\frac{4}{1-\beta^2}$   & \cref{eq:q-to-t}  &                           &                               \\
        $\beta$    &   $\mathbb H_\geq$    &   $\Jbub{n}$, $\bar E_{5,6}$       &   $\sqrt{1-4/t}$  &   $\beta$                 &                   &                           &                               \\
        $q$        &   $\mathbb D$         &   $E_{1,2,3}$ ($t\in\mathbb C$)    &                   &                           &   $q$             &   $e^{2\pi i\tau}$        &                               \\
        $\tau$     &   $\mathbb H_>$       &   $E_{1,2,3}$ ($t\in\mathbb C$)    &   \cref{eq:t-to-tau}  &                       &   $\frac{\log q}{2\pi i}$   &   $\tau$        &   \cref{eq:theta}             \\
        $\theta$   &   $\mathbb R$         &   $E_{1,2,3}$ ($t\in\mathbb R$)    &   $*$             &                           &                   &                           &   $\theta$                    \\
        \bottomrule
    \end{tabular}
    \caption{Summary of the different variables.
            For the domains, $\mathbb H$ denotes the upper half-plane (including the real line for $\mathbb H_\geq$ and excluding it for $\mathbb H_>$) and $\mathbb D$ is the open unit disk.
            The ``Describes'' column indicates which quantities are most naturally evaluated in terms of this variable.
            The remaining columns show how to obtain the variable via the mappings $\map{x}{y}$.
            Missing entries are filled in through composition: $\map{x}{z}=\map{x}{y}\circ\map{y}{z}$.
            The mappings $\map{x}{\theta}$ are typically not defined since $\map{\theta}{y}$ is not surjective, but $\map{t}{\theta}$ (marked $*$) is described for $x\in\mathbb R$ in \cref{sec:practical}.
    }
    \label{tab:map}
\end{table}

\section{The evaluation of $E_5$ and $E_6$ in four dimensions}\label{sec:E5E6}

The master integral $\bar E_5$ satisfies the differential equation~\mainref[eq.~(C.20)]
\begin{equation}\label{eq:diffeq-E5bar}
    (t-4)\bigg[\frac{(t-4)t^2}{4}\frac{\p^2}{\p t^2} + \frac{(t-3)t}{2}\frac{\p}{\p t} + \frac{1}{2}\bigg]\bar E_5(4;t)
        = \beta^2\bigg[\frac{\p^2}{\p\beta^2} - \frac{2}{\beta^2-1}\bigg]\bar E_5(4;\beta) 
        = \mathscr S_5(\beta)\,,
\end{equation}
where the change of variables $t=4/(1-\beta^2)$ greatly simplifies the equation.
The source term is~\mainref[eqs.~(C.21-24)]
\begin{gather}\label{eq:S5}
    \begin{aligned}
        \mathscr S_5(\beta) &= \mathscr S_\rat(\beta) + \mathscr S_J(\beta) + \mathscr S_{E_1}(\beta) \,,\\
        \mathscr S_\rat(\beta) &\coloneq
            \frac{68 \beta^{4} - 44 \beta^{2} + 24 - \beta^{2} \big(\beta^{2}+5\big) \pi^{2}}{6(\beta^2 -1)^{2}}
            + \frac{2 \beta^{2} \zeta(3)}{3(\beta^2 -1)}\,,\\
        \mathscr S_J(\beta) &\coloneq 
            \frac{\beta^{2}}{3(\beta^{2}-1)} \Jbub{3}(\beta) 
            + \frac{2}{\beta^2-1}\Jbub{1}(\beta)\Jbub{2}(\beta)
            - \frac{\beta^{4}+5}{(\beta^2-1)^2}\Jbub{2}(\beta)
            - \frac{12}{(\beta^{2}-1)^2} \Jbub{1}(\beta)^{2} \eqnbreak\qquad
            + \frac{(\pi^2-4)\beta^4+(8-\pi^2)\beta^2-52}{2(\beta^{2}-1)^2}\Jbub{1}(\beta)\,,\\
        \mathscr S_{E_1} &\coloneq\bigg[
              s_2(\beta)\frac{\p^2}{\p\beta^2}
            + s_1(\beta)\frac{\p}{\p\beta}
            + s_0(\beta)\bigg]E_1(2;\beta)\,,
    \end{aligned}
    \\
    s_0(\beta) = \frac23\,  \frac{\beta^2(2\beta^2-1)}{(\beta^2-1)^2}\,,\quad
    s_1(\beta) = \frac{2\beta}{3}\frac{3\beta^4-7\beta^2+3}{\beta^2-1}\,,\quad
    s_2(\beta) = \frac{\beta^2(4\beta^2-3)}{2}\,.
\end{gather}
This has the Wrońskian solution~\mainref[eq. (C.26)]
\begin{equation}\label{eq:E5-generic}
    \bar E_5(4;\beta) = 
        c_1\, g_1(\beta) + c_2\, g_2(\beta)
        - g_1(\beta)\int_{\xi_1}^{\beta} \mathscr S_5(\xi) g_2(\xi) \frac{\d \xi}{\xi^2}
        + g_2(\beta)\int_{\xi_2}^{\beta} \mathscr S_5(\xi) g_1(\xi) \frac{\d \xi}{\xi^2}\,,
\end{equation}
where
\begin{equation}\label{eq:g1-g2-s2}
    g_1(\beta) = \beta^2-1\,, \qquad
    g_2(\beta) = \frac{\beta^2 - 1}{4}\log\bigg(\frac{\beta+1}{\beta-1}\bigg)-\frac{\beta}{2}
\end{equation}
are the homogeneous solutions, $c_1$ and $c_2$ are constants to be determined in \cref{sec:c1c2}, and the integration limits $\xi_1$ and $\xi_2$ may be set arbitrarily by adjusting $c_1$ and $c_2$.
After performing integration by parts to avoid derivatives of $E_1(2;\beta)$ in the integrand,~\mainref[eq.~(C.31)]
\begin{equation}
    \label{eq:E5-main}
    \bar E_5(4;\beta) 
        = c_1g_1(\beta) + c_2g_2(\beta) + \frac{s_2(\beta)}{\beta^2}E_1(2;\beta) + \I\big[\beta; g_1(\beta),g_2(\beta)\big]\,,
\end{equation}
where, for convenience, we express the integrals as
\begin{equation}\label{eq:E5-I}
    \begin{aligned}
        \I[\beta; G_1,G_2] =
            &-G_1\int_{\xi_1}^{\beta} \bigg[\frac{\mathscr S_J(\xi) + \mathscr S_\rat(\xi)}{\xi^2}g_2(\xi) + E_1(2;\xi)h_2(\xi) \bigg] \d\xi\\
            &+G_2\int_{\xi_2}^{\beta} \bigg[\frac{\mathscr S_J(\xi) + \mathscr S_\rat(\xi)}{\xi^2}g_1(\xi) + E_1(2;\xi)h_1(\xi) \bigg] \d\xi
    \end{aligned}
\end{equation}
and the integration by parts has yielded
\begin{equation}\label{eq:E5-h}
    h_n(\xi) = 
          \frac{\p^2}{\p \xi^2}\frac{s_2(\xi)g_n(\xi)}{\xi^2}
        - \frac{\p}{\p \xi}\frac{s_1(\xi)g_n(\xi)}{\xi^2}
        + \frac{s_0(\xi)g_n(\xi)}{\xi^2}\,,
\end{equation}
that is,
\begin{align}
    \label{eq:h1-h2}
    h_1(\xi) &= \frac{54 \xi^6-57 \xi^4+11 \xi^2-6}{3\xi^2(\xi^2-1)}\,,\\
    h_2(\xi) &= \frac{ (\xi^2 -1) \big(54 \xi^{6}-57 \xi^{4}+11  \xi^{2}-6\big)\log\Big(\frac{\xi+1}{\xi-1}\Big)
        -2 \xi  \big(54 \xi^{6}-93 \xi^{4}+31 \xi^{2}+6\big)
        }{12(\xi^2-1)^2 \xi^2}\,.\notag
\end{align}
The master integral $E_6$ can be expressed in terms of derivatives of $E_1$ and $E_5$~\mainref[eq.~(C.34)],
\begin{align}\label{eq:E6-main}
    \bar E_6(4;t)
          &= 
        \bigg[
            \frac{(t-16)(t-4)}{12t^2}\Big(t\frac{\p}{\p t }\Big)^2
            + \frac{t-10}{12t} t\frac{\p}{\p  t}
            + \frac{1}{24}
            \bigg] E_1(2;t)
        - \bigg[
            \frac{\p}{\p t} + \frac{1}{t}
            \bigg] \bar E_5(4;t)\eqnbreak
        + \frac{-2\Jbub3(t) + 6\Jbub2(t) + 3(4 - \pi^2)\Jbub1(t)}{24t}
        - \frac{\zeta(3) + 5}{3t} + \frac{\pi^2}{12t}\,,
\end{align}
where it follows from $\frac{\p}{\p t}=\frac{(\beta^2-1)^2}{8\beta}\frac{\p}{\p\beta}$ and \cref{eq:E5-main} that%
\footnote{
    Note that the original solution, \cref{eq:E5-generic}, produces a sum of integrals similar to our $\I$, but which, if used as-is, would give the property $\frac{\p}{\p\beta}\I[\beta;g_1(\beta),g_2(\beta)] = \I[\beta;g'_1(\beta),g'_2(\beta)]$.
    This does not hold here due to the integration by parts performed to eliminate derivatives of $E_1$ in the integrand, resulting in the $\frac{s_2(\beta)}{\beta^2}E_1(\beta)$ term in \cref{eq:E5-main} and the combination of $E_1$ and $\frac{\p}{\p t}E_1$ in \cref{eq:dE5dt}.
    The latter may be derived by noting that $\smash{\frac{t-16}{2t}}=\smash{\frac{s_2(\beta)}{\beta^2}}$ and that $\smash{\frac{t^{2}-28 t +48}{3 t^2 (t-4)}}$ is equal to either of 
    \begin{equation*}
        \frac{(\beta^2-1)^2}{8\beta}\bigg[\frac{\p}{\p\beta}\frac{s_2(\beta)}{\beta^2} - g_1(\beta)h_2(\beta) + g_2(\beta)h_1(\beta)\bigg]
        = \frac{(\beta^2-1)^2}{8\beta}\big[g_2(\beta)g'_1(\beta) - g'_2(\beta)g_1(\beta)\big] \bigg[s_1(\beta) + \frac{\p}{\p\beta}\frac{s_2(\beta)}{\beta^2}\bigg]\,,
    \end{equation*}
    where the second expression comes from applying the $t$-derivative \emph{before} doing integration by parts.}
\begin{align}\label{eq:dE5dt}
    \frac{\p\bar E_5(4;t)}{\p t}
        &= 
         \bigg[\frac{t-16}{2t}\frac{\p}{\p t} + \frac{t^2-28 t +48}{3 t^2 (t-4)}\bigg]E_1(2;t)\eqnbreak\qquad
        + \frac{(\beta^2-1)^2}{8\beta}\Big(2\beta c_1 + \tfrac12\Jbub1(\beta)c_2 + \I\big[\beta;2\beta, \tfrac12\Jbub1(\beta)\big]\Big)\,,
\end{align}
The remainder of this section is dedicated to the numerically practical evaluation of $\I[\beta;G_1,G_2]$, and (in \cref{sec:c1c2}) to the determination of the constants $c_1$ and $c_2$ in \cref{eq:E5-main}.

\subsection{The spacelike region}\label{sec:spacelike}

As shown in \cref{fig:tau,fig:q,fig:tsun}, the real $t$-line naturally decomposes into four regions, of which we now consider the spacelike one, i.e., $t<0$ or $\beta>1$.
We split the integrand of \cref{eq:E5-I} into multiple pieces, some of which yield integrals that can be evaluated analytically and others that require numerical evaluation.
Importantly, we are free to choose different integration limits $\xi_1,\xi_2$ separately for each piece.
We will treat the analytic pieces as indefinite integrals, leaving $\xi_1,\xi_2$ entirely unspecified and compensating with our choice of $c_1,c_2$.
For the integrals evaluated numerically, we choose \mbox{$\xi_1=\xi_2=\infty$}, ensuring that the integrals converge, since this avoids dependence on arbitrarily chosen finite limits and allows use of the known $t\to0$ behavior of $\bar E_5$, namely \cref{eq:elliptics-t0}. Thus:

\bulletpoint
The contribution from $\mathscr S_\rat$ can be evaluated analytically, and we leave it as an indefinite integral.%
\footnote{
    In fact, since $\mathscr S_\rat(\beta)g_1(\beta)/\beta^2$ does not vanish as $\beta\to\infty$, it would be inconvenient to write it as a definite integral, since the only conveniently accessible limit, $\xi_2=\infty$, is excluded.
    Below, we encounter (and solve) the same problem for the contributions involving $E_1$, but we can avoid it altogether thanks to the integral's analytic solvability.}
Writing $\Irat{n}(\beta)\coloneq \int \frac{\mathscr S_\rat(\beta) g_n(\beta)}{\beta^2} \d\beta$, we choose the primitive functions
\begin{align}\label{eq:Irat}
    \Irat{1}(\beta) 
        &= \frac{\pi^2-8}{2} \logpm{\beta}
            + \frac{24 + \big[68 + 4\zeta(3) - \pi^2\big]\beta^2}{6\beta}\,,\\
    \Irat{2}(\beta) 
        &= \bigg[\frac{\pi^2-8}{16}\logpm{\beta}
                + \frac{24 + \big[68 + 4\zeta(3) - \pi^2\big]\beta^2}{24\beta}\bigg]\logpm{\beta}
            \eqnbreak\quad
            + 2\logsq{\beta}
            + \frac{\zeta(3)}{3} 
            + \frac{8 - \pi^2}{4(\beta^2-1)}
            + \frac{68 - \pi^2 + 4\zeta(3)}{12}\,,
\end{align}
with constants of integration chosen so that their contribution to $\bar E_5(4;\beta)$ is finite as $\beta\to\infty$ [see \cref{eq:I-asymptotic:rat}].

\bulletpoint
The contribution stemming from $\mathscr S_J$ can be evaluated analytically as well: in terms of polylogarithm weight, $\mathscr S_J(\xi)$ has weight $3$ and $g_2(\xi)$ has weight $1$, so the integral of their product has weight $5$. It is easily obtained using \texttt{hyperInt}~\cite{Panzer:2014caa}.
However, the resulting expression is very long and involves generalized polylogarithms, the evaluation of which is no faster than evaluating the integral numerically.
Therefore, we do exactly this, numerically evaluating the convergent integrals
\begin{equation}\label{eq:IJ}
    \IJ{n}(\beta)\coloneq -\int_\beta^\infty \frac{\mathscr S_J(\xi) g_n(\xi)}{\xi^2} \d\xi\,,
\end{equation}
where we have set $\xi_n=\infty$ and interchanged the limits.

\bulletpoint
The contributions involving $E_1(2;\beta)$ must be evaluated numerically.
From \cref{eq:h1-h2}, as $\xi\to\infty$, 
\begin{equation}
    h_1(\xi) = 18\xi^2 - 1 + \O(1/\xi^2)\,,\qquad
    h_2(\xi) = \frac{9}{5\xi^3}+\O(1/\xi^5)\,,
\end{equation}
and from \cref{eq:elliptics-t0},
\begin{equation}\label{eq:E1-limit}
    E_1(2;\xi)= -7\zeta(3) + \frac{7\zeta(3) - 6}{4\xi^2} + \O(1/\xi^4)\,.
\end{equation} 
Therefore, $\int_\beta^\infty h_2(\xi) E_1(2;\xi)$ converges but $\int_\beta^\infty h_1(\xi) E_1(2;\xi)$ does not.
To remedy this, we subtract the leading large-$\xi$ behavior of $E_1(2;\xi)$, writing $h_1(\xi) E_1(2;\xi) = \mathscr H_\Reg(\xi) + \mathscr H_\Div(\xi)$ with
\begin{equation}\label{eq:Hdiv}
  \mathscr H_\Div(\xi) \coloneq h_1(\xi) \bigg(-7\zeta(3)+\frac{7\zeta(3) - 6}{4\xi^2}\bigg)\,,\qquad
  \mathscr H_\Reg(\xi) \coloneq h_1(\xi) E_1(2;\xi) - \mathscr H_\Div(\xi)\,.
\end{equation}
Now, the divergent integral can be evaluated analytically,
\begin{align}\label{eq:Idiv}
    \I_\Div(\beta) \coloneq \int \mathscr H_\Div(\beta)\d\beta
        & = -42\zeta(3)\beta^3 
        + {77\zeta(3)-54\over2}\beta
        + \frac{203\zeta(3) - 30}{12\beta}
        + \frac{6-7\zeta(3)}{6\beta^3} \eqnbreak\quad
        + \frac{7\zeta(3) + 2}{4}\logpm{\beta}\,,
\end{align}
leaving the convergent numerical integrals
\begin{equation}\label{eq:IE}
    \IE{1}(\beta) \coloneq -\int_\beta^\infty \mathscr H_\Reg(\xi)\d\xi\,,
    \qquad
    \IE{2}(\beta) \coloneq -\int_\beta^\infty h_2(\xi) E_1(2;\xi)\,.
\end{equation}
Summing up the contributions from \cref{eq:Irat,eq:IJ,eq:Idiv,eq:IE}, we have
\begin{align}\label{eq:E5-subthr}
    \I[\beta; G_1,G_2]
        &= G_2\Big[\Irat{1}(\beta) + \IJ{1}(\beta) + \IE{1}(\beta) + \Idiv(\beta) \Big]\notag\\
        &- G_1\Big[\Irat{2}(\beta) + \IJ{2}(\beta) + \IE{2}(\beta) \Big]\,,
\end{align}
which can be plugged into \cref{eq:E5-main}.

\subsection{Constants of integration}\label{sec:c1c2}
We can now determine $c_1,c_2$ in \cref{eq:E5-main} by comparing \cref{eq:E5-subthr} with the asymptotic behavior as $t$ approaches $0$ from below (i.e., as $\beta\to\infty$).
We can, in fact, directly see that $c_2=0$: all of $\bar E_5(4;\beta)$, $g_1(\beta)$ and $\I[\beta;g_1(\beta),g_2(\beta)]$ are even under $\beta\to-\beta$, but $g_2(\beta)$ is odd.
To determine $c_1$, we use the expansion $g_2(\beta) = -\frac{1}{3\beta} - \frac{1}{15\beta^3} + \O(1/\beta^5)$ along with \cref{eq:Irat,eq:IJ,eq:Idiv,eq:IE}, giving
\begin{subequations}\label{eq:I-asymptotic}
    \begin{alignat}{3}
        g_2(\beta)\Irat{1}(\beta) - g_1(\beta)\Irat{2}(\beta) 
            &= \frac{\pi^{2} - 4\zeta(3) - 68}{12}
            &&+\frac{4-\pi^2}{4\beta^{2}} 
            && +\O\Big(\frac{1}{\beta^4}\Big)\,,
            \label{eq:I-asymptotic:rat}\\
        g_2(\beta)\IJ{1}(\beta) - g_1(\beta)\IJ{2}(\beta) 
            &= 
            &&\phantom{+}\; \frac{\pi^2 - 4}{12\beta^2}
            && + \O\Big(\frac{1}{\beta^4}\Big)\,,
            \label{eq:I-asymptotic:J}\\
        g_2(\beta)\IE{1}(\beta) - g_1(\beta)\IE{2}(\beta) 
            &= - \frac{63\zeta(3)}{10}
            && + \frac{4698 - 1147\zeta3}{1440\beta^2}
            && + \O\Big(\frac{1}{\beta^4}\Big)\,,
            \label{eq:I-asymptotic:E}\\
        \frac{s_2(\beta)}{\beta^2} E_1(2;\beta) + g_2(\beta)\I_\Div(\beta)
            &= \frac{180+119\zeta(3)}{30}
            && + \frac{4662 - 12713\zeta(3)}{1440\beta^2}
            && + \O\Big(\frac{1}{\beta^4}\Big)\,,
            \label{eq:I-asymptotic:div}
    \end{alignat}
\end{subequations}
where we have cancelled the divergent part of \mbox{$g_2(\beta)\I_\Div(\beta)=14\zeta(3)\beta^2 + \O(1)$} against that of the boundary term in \cref{eq:E5-main}.%
\footnote{
    Note that the total $E_1$ contribution is the simpler-looking
    \begin{equation*}
        \frac{s_2(\beta)}{\beta^2} E_1(2;\beta) + g_2(\beta)\IE{1}(\beta) - g_1(\beta)\IE{2}(\beta) - g_1(\beta)\I_\Div(\beta) = \frac{18-7\zeta(3)}{3} + \frac{52-77\zeta(3)}{8\beta^2} + \O\big(1/\beta^4\big)\,.
    \end{equation*}
    This would appear more directly if the original expression~\mainref[eq. (C.26)] were used, but this complication is more than made up for by having only one elliptic function---namely $E_1(2;\xi)$, not its derivatives---in the integrand.}
The absence of a $\beta^2$ term in \cref{eq:I-asymptotic} immediately tells us that $c_1=0$, and summing up the contributions, we see that the total asymptotic behavior precisely matches the large-$\beta$ expansions of $\bar E_5(4;\beta)$ from \cref{eq:elliptics-t0}:
\begin{equation}\label{eq:E5-asymptotic}
    \bar E_5(4;\beta) 
        = \frac{4 + \pi^2 - 32\zeta(3)}{12}
        + \frac{105\zeta(3) - 8\pi^2 - 34}{48\beta^2}
        + \O\big(1/\beta^4\big)\,.
\end{equation}
This equality holds to all orders in $1/\beta^2$, as is easily verified by extending the expansions above.
Since $c_1=c_2=0$, we will ignore the constants of integration from now on.

\subsection{The timelike region, and everywhere else}\label{sec:timelike}

\Cref{eq:E5-subthr} holds for $\beta>1$ (i.e., $t<0$), but in fact it remains valid when $0<t<4$, corresponding to $\beta=i\tilde\beta$ with $\tilde\beta>0$.
This is seen by changing the integration variable to $\tilde\beta$ (and, similarly, to $\tilde\xi$ with $\xi=i\tilde\xi$).
In the cases relevant for $\bar E_5$ and $\bar E_6$,
$G_1$ and the integrals multiplying it stay real, and $G_2$ and those multiplying it become purely imaginary, as follows from \mbox{$\log[(i\tilde\xi+1)/(i\tilde\xi-1)] = -2i\tan^{-1}(1/\tilde\xi)$} etc. All considerations about convergence carry over since $\beta=\infty$ and $\tilde\beta=\infty$ both correspond to $t=0$.
Thus, \cref{eq:E5-main} holds for all $t$ below the two-pion threshold at $t=4$.

However, we now demonstrate another method, which is equivalent to the above for $t<4$, but which also generalizes to the above-threshold region $t>4$, and indeed to the entire complex plane.

The above-threshold region ($t>4$) corresponds to $\beta\in(0,1)$, but here there is a branch cut along the real line requiring that $t$, and therefore $\beta$, be given a small imaginary part.%
\footnote{
    Note that, for the purposes of $\I[\beta;G_1,G_2]$, there is nothing special happening at the four-pion threshold ($t=16$, $\beta=\sqrt3/2$).
    Any new behavior associated with the opening of the four-pion cut is inherited directly from $E_1(2;t)$ in the integrand.}
There is a singularity at $\beta=1$, so there is no longer convenient access along the real line to any well-behaved limit.
Instead, we piggy-back off the spacelike solution by routing the numerical integrals along complex contours leading to an arbitrary point $\beta_\star>1$, from which they can continue to $\infty$ along the real line.
Thus, for some arbitrary contour $\mathcal C$ that goes from $\beta$ to $\beta_\star$ without crossing the cut,
\begin{align}\label{eq:E5-everywhere}
    \I[\beta; G_1,G_2]
        &= G_2\bigg[\Irat{1}(\beta) + \IJ{1}(\beta_\star) + \IE{1}(\beta_\star) + \Idiv(\beta) - \int_{\mathcal C} \bigg(\frac{\mathscr S_J(\xi)g_1(\xi)}{\xi^2} + \mathscr H_\Reg(\xi)\bigg)\d\xi \bigg]\notag\\
        &- G_1\bigg[\Irat{2}(\beta) + \IJ{2}(\beta_\star) + \IE{2}(\beta_\star) - \int_{\mathcal C} \bigg(\frac{\mathscr S_J(\xi)g_1(\xi)}{\xi^2} + h_2(\xi)E_1(2;\xi)\bigg)\d\xi \bigg],
\end{align}
which is valid for all $t\in\mathbb C$ except for the cut along $[4,\infty)$.

\subsection{Practical considerations}\label{sec:practical}

There are numerous pitfalls in the method described above, so here we provide further details that are more closely aligned with how $\bar E_5$ and $\bar E_6$ are implemented in practice.

\bulletpoint
Nothing about \cref{eq:E5-everywhere} is specific to $\bar E_5$, so the same integrals can be used for $\bar E_6$, further aided by the linearity of $\I[\beta;G_1,G_2]$ in $G_1,G_2$.
Schematically,
\begin{equation}\label{eq:Ebar-schematic}
    \bar E_n(4;t) = B_n + \I\big[\beta; G_{1n}, G_{2n}\big]\,,
\end{equation}
where
\begin{equation}
    \begin{gathered}
        G_{15} = g_1(\beta)\,,\qquad
        G_{25} = g_2(\beta)\,,\qquad
        B_5 = \frac{s_2(\beta)}{\beta^2}E_1(2;\beta)\,,\\
        G_{16} = -\frac{4}{t^2} - \frac{g_1(\beta)}{t} = 0\,,\qquad
        G_{26} = -\frac{\Jbub1(\beta)}{4t^2\beta} - \frac{g_2(\beta)}{t}\,,\label{eq:Gdef}
    \end{gathered}
\end{equation}
and $B_6$ is the combination of $\Jbub{n}$ and (derivatives of) $E_1$ appearing in \cref{eq:E6-main}, minus that in \cref{eq:dE5dt}, minus $B_5/t$. 

\bulletpoint
\Cref{eq:E5-everywhere} has the added benefit that $\beta_\star$ can be held constant, meaning that many of the enclosed integrals need to be evaluated only once.
Partly for aesthetic reasons, we choose $\beta_\star=\sqrt3$, corresponding to $t_\star=\map{\beta}{t}(\sqrt3)=-2$ and $\tau_\star=\map{t}{\tau}(-2)=i/\sqrt{12}$, making it a rarity for which all of $\tau$, $t$ and $\beta$ are algebraic.%
\footnote{
    Other such points include $(t=-8, \tau=i/\sqrt6, \beta=\sqrt{3/2})$ and $(t=-32, \tau=i/\sqrt3, \beta=3/\sqrt8)$.}
Evaluating the integrals using much higher working precision than what is otherwise needed, we obtain
\begin{equation}\label{eq:star-values}
    \begin{aligned}
    \IJ1(\beta_\star) &= -0.897\,428\,059\,322\,249\,442\,551\,141\ldots\,,\\
    \IJ2(\beta_\star) &= +0.007\,806\,884\,759\,652\,421\,539\,440\ldots\,,\\
    \IE1(\beta_\star) &= -6.114\,617\,584\,976\,858\,497\,987\,754\ldots\,,\\
    \IE2(\beta_\star) &= +2.827\,177\,949\,642\,984\,743\,645\,142\ldots\,.
    \end{aligned}
\end{equation}
These values are accurate to many more digits than shown here.

\bulletpoint
A problem with $\IJ{n}$ and $\IE{n}$ [recall \cref{eq:IJ,eq:IE}] is that, while the integrals are theoretically convergent, numerical noise in the large-$\xi$ limit of the integrands typically spoils that convergence in practice.
To remedy this, we perform the rearrangement (written here for a generic integrand $f$) 
\begin{equation}\label{eq:hybrid-integral}
    \int_\beta^\infty f(\xi)\d\xi 
        = \int_\beta^\chi f(\xi)\d\xi + \int_\chi^\infty \sum_{n=2}^\infty \frac{a_n}{\xi^n} \d\xi
        = \int_\beta^\chi f(\xi)\d\xi + \sum_{n=2}^\infty \frac{a_n}{(1-n)\chi^{n-1}}\,,
\end{equation}
where $f(\xi)=\sum_{n=2}^\infty \frac{a_n}{\xi^n}$ when expanded around $\xi=\infty$.
We choose the cutoff $\chi$ small enough that $\int_\beta^\chi f(\xi)\d\xi$ can be evaluated numerically without noise issues, but large enough that the sum can be truncated, without loss of precision, at some finite $n=N$ such that $\{a_n\}_{n=2}^N$ are known.
We estimate the truncation error based on the relative difference between the exact and series-expanded integral at the cutoff, resulting in the uncertainty
\begin{equation}\label{eq:hybrid-error}
    \left|\frac{f(\chi) - \sum_{n=2}^N\frac{a_n}{\chi^n}}{f(\chi)}\right| \sum_{n=2}^N \frac{a_n}{(1-n)\chi^{n-1}}\,,
\end{equation}
which we add in quadrature to the numerical uncertainty of $\int_\beta^\chi f(\xi)\d\xi$. 
The optimal choice of $\chi$ depends on the working precision of the numerics and the available $N$. In our case, we have found $\chi=10$ to be a good choice for $\IE{n}$, and $\chi=100$ for $\IJ{n}$.
The Taylor expansions are given in \cref{app:t0-series}.

A related issue is that the numerical integration method we use (\texttt{quad} from the \texttt{mpmath} library~\cite{mpmath}) does not account for the uncertainty of the integrand.
In lieu of proper error propagation, we may estimate the integrand-derived error of $\int_\beta^\chi f(\xi)\d\xi$ as $\int_\beta^\chi \delta f(\xi)\d\xi$, where $\delta f$ is the standard error of $f$.
With our choice of $\chi$, this contribution to the error is mostly negligible, so we normally omit it to speed up the computations.
This also holds for the numerically evaluated contour integrals.

\bulletpoint
For $\IE{n}$, which is the numerically most expensive part of the calculation, a significant bottleneck is the conversion $\map{\beta}{\tau}(\xi)$, which is necessary for evaluating $E_1(2;\xi)$ and which requires the complex machinery of \cref{sec:t-tau-q}.
A more efficient approach is to perform the integral in the $\tau$-plane, where $E_1$ can be evaluated directly: the function $\map{\tau}{\beta}$ needed to obtain $h_n(\xi)$ is significantly cheaper than $\map{\beta}{\tau}$, as is the Jacobian
\begin{equation}
    \map{\tau}{\beta}'(\tau) = -\frac{\pi i}{\beta t} \big[1+6 \E(0,2,\psi_t; q)\big] = \frac{\pi i}{\beta t}\frac{
            \big[\eta(q)\eta(q^2)\big]^4 + 9\big[\eta(q^3)\eta(q^6)\big]^4
        }{
            \eta(q)\eta(q^2)\eta(q^3)\eta(q^6)
        }\,,
\end{equation}
where $q=e^{2\pi i\tau}$ and $ t=\map{\tau}{t}(\tau)=\frac{4}{1-\beta^2}$, and the two variants follow from \cref{eq:Dqt-eisen} and~\eqref{eq:Dqt-eta}, respectively.

For contour integrals, we replace $\int_{\mathcal C} h_n(\beta)E_1(2;\xi)\d\xi$ by
\begin{equation}
    \int_{\mathcal C_\tau(\beta)} h_n\big(\map{\tau}{\beta}(\tau)\big) E_1(2;\tau) \map{\tau}{\beta}'(\tau) \d\tau\,,
\end{equation}
where $\mathcal C_\tau(\beta)$ is the straight-line contour from $\map{\beta}{\tau}(\beta)$ to $\tau_\star = i/\sqrt{12}$.
Note that a nice property of this particular $\tau_\star$ is that it is connected to any $\tau\in\map{t}{\tau}(\mathbb C)$ by a straight line entirely contained in $\map{t}{\tau}(\mathbb C)$.%
\footnote{
    The image $\map{t}{\tau}(\mathbb H)$, where $\mathbb H$ is the upper half-plane, is the shaded portion of \cref{fig:tau} (plus the extension to $\Im\tau=\infty$).
    Its boundary is $\map{\theta}{\tau}(\mathbb R)$.
    Likewise, $\map{t}{q}(\mathbb H)$ is the shaded portion of \cref{fig:q} (shown in its entirety) and is bounded by $\map{\theta}{q}(\mathbb R)$.
    The images of the lower half-plane are obtained through $\tau\to-\tau^*$ or $q\to q^*$.
}

\begin{figure}[p!]
    \newcommand{\ex}{\text{ex}}
    \hspace{-.5cm}
    \includegraphics{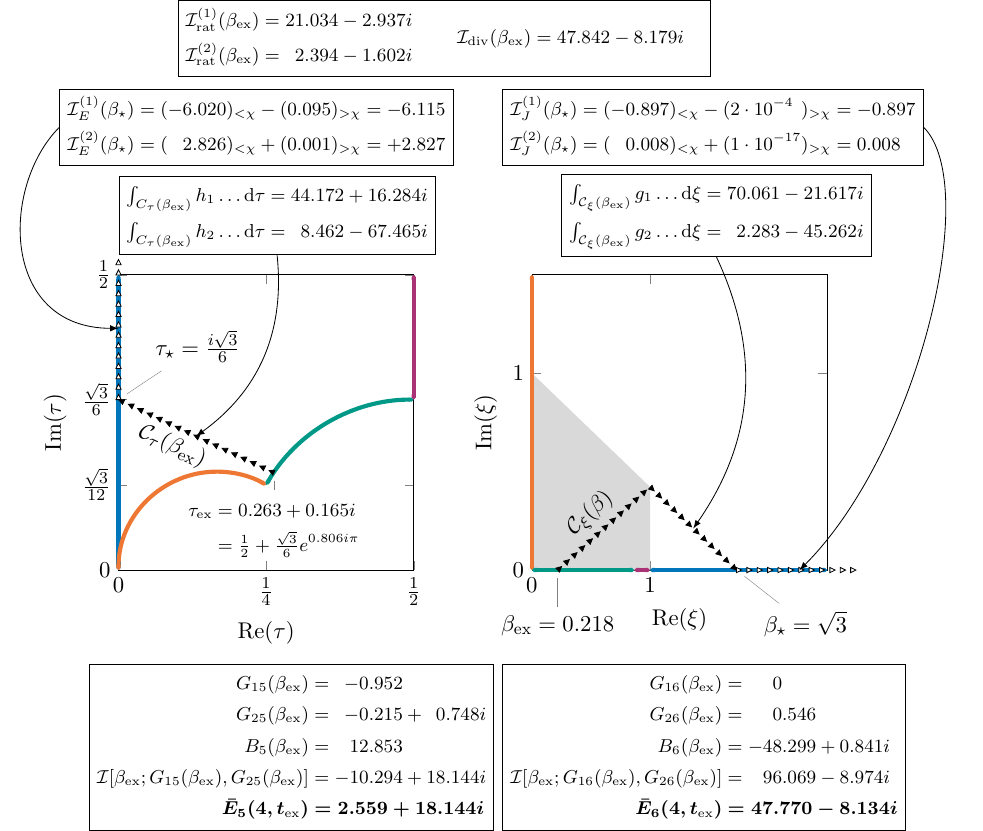}
    \fignumericexamplecaption
    \label{fig:numeric-example}
\end{figure}
\afterpage{\clearpage} 

For integrals along the real line, such as $\IE{n}(\beta_\star)$, we run into the problem that $\map{t}{\tau}$, as given in \cref{sec:t-tau-q}, breaks down for real $t$.
Instead, we turn to the $\theta$-parametrization of \cref{eq:theta}, which allows $\int_\beta^\infty h_2(\xi) E_1(2;\xi)\d\xi$ to be replaced by
\begin{equation}
    \int_{\map{\beta}{\theta}(\beta)}^\infty h_n\big(\map{\theta}{\beta}(\theta)\big) E_1\big(2;\map{\theta}{\tau}(\theta)\big) \map{\theta}{\beta}'(\theta) \d\theta\,,
\end{equation}
and similarly for $\int_\beta^\infty \mathscr H_\Reg(\xi) \d\xi$.
While $\map{\theta}{\beta}$ and $\map{\theta}{\beta}'$ are straightforward combinations of known functions, this is not the case for the inverse map $\map{\beta}{\theta}$.
However, it only needs to be applied twice---once to the integration limit, and once to $\chi$ when using \cref{eq:hybrid-integral}---so a crude method is acceptable here.
Extrapolating $\map{\beta}{\tau}(\beta+i\epsilon)$ to $\epsilon=0$ and then inverting \cref{eq:theta} is an option, but we use the simpler procedure of applying the Newton-Raphson method to the initial estimate%
\footnote{
    This is informed by asymptotic approximations of $\map{\theta}{t}(\theta)$ for $\theta\to\pm\infty$, with some highly heuristic interpolation thereof to make the small-$t$ region work.
    $\map{\theta}{t}'(\theta)$ goes as $e^{-1/|\theta|}$ near $\theta\sim t\sim 0$, which poses a problem for convergence. Still, thanks to the availability of series expansions, there is no need to compute values very close to zero in this way.}
\begin{equation}
    \theta_\text{est.} = 
    \begin{cases}
        -\tfrac{8}{\pi}\log(-t)         &   \text{if } t < -10\,,\\
        \min(t,-3)                      &   \text{if } -10 \leq t < 0\,,\\
        \max(t,+3)                      &   \text{if } 0\leq t < 20\,,\\
        -\tfrac{48}{\pi\sqrt3}\log(+t)  &   \text{if } 20 \leq t\,,
    \end{cases}
\end{equation}
which rapidly converges to a $\theta$ such that $\map{\theta}{t}(\theta)=t$.

\bulletpoint
For integrals not involving $E_1$, we keep the integration contour in the $\xi$-plane.
Sufficiently far below threshold, a straight line from $\beta$ to $\beta_\star=\sqrt3$ works. However,  when $\Im\beta$ is small, the line approaches the branch cut on $[0,1]$ too closely, and we must reroute it.
Arbitrarily and conservatively defining ``too close to the cut'' as ``intersecting the region where $|\Re(\beta)|<1$ and $|\Im(\beta)| + \frac{|\Re(\beta)|}{|\Re(\beta_\star)|}<1$'', we route the contour via the region's corner point, as illustrated in \cref{fig:numeric-example}.
We denote this piecewise straight contour by $\mathcal C_\xi(\beta)$.

\bulletpoint
With the contours in hand, a more detailed version of \cref{eq:E5-everywhere} is
\begin{align}\label{eq:Ebar-detailed}
    \I[\beta; G_1,G_2]
        &= G_2\Big[\Irat{1}(\beta) + \IJ{1}(\beta_\star) + \IE{1}(\beta_\star) + \Idiv(\beta)\Big]\notag\\
        &- G_1\Big[\Irat{2}(\beta) + \IJ{2}(\beta_\star) + \IE{2}(\beta_\star)\Big]\notag\\
        &- G_2\bigg[
            \int_{\mathcal C_\xi(\beta)}\Big(\frac{\mathscr S_J(\xi)g_1(\xi)}{\xi^2} - \mathscr H_\Div(\xi)\Big)\d\xi
            \notag\\&\qquad\qquad
            + \int_{\mathcal C_\tau(\beta)} h_1\big(\map{\tau}{\beta}(\tau)\big) E_1(2;\tau) \map{\tau}{\beta}'(\tau) \d\tau
            \bigg] \notag\\
        &+ G_1\bigg[
            \int_{\mathcal C_\xi(\beta)}\frac{\mathscr S_J(\xi)g_2(\xi)}{\xi^2}\d\xi
            + \int_{\mathcal C_\tau(\beta)} h_2\big(\map{\tau}{\beta}(\tau)\big) E_1(2;\tau) \map{\tau}{\beta}'(\tau) \d\tau
            \bigg] \,.
\end{align}
Note that we have split $\mathscr H_\Reg$ into its constituents according to \cref{eq:Hdiv}, since convergence is no issue here.
An illustrated overview of this equation at a specific $\beta$ is given in \cref{fig:numeric-example}.

\section{Results}\label{sec:results}

Having derived means for computing the master integrals in the preceding sections, we are now ready to show the results and analyse the precision and performance of the various methods.
To recapitulate, the full set of methods is the following:
\begin{itemize}
    \item \emph{Bessel integrals} [\cref{eq:E1-bessel,eq:E2-bessel,eq:E3-bessel}] allow for the computation of $E_{1,2,3}(2;t)$ for \mbox{$t\in(-\infty,16)$}.
    Via \cref{eq:E4-finite}, this also gives $\bar E_4(4;t)$.
    \item \emph{Elliptic polylogarithms} [\cref{sec:polylog}]
    allow the same integrals to be computed for all $t$, granted that that $t$ can be translated to $q$ via the methods of \cref{sec:t-tau-q}.
    \item \emph{Eisenstein series} [\cref{sec:eisenstein}]
    is a variant of the elliptic polylogarithm method, aiming for greater numerical efficiency.
    \item \emph{Poisson resummation} [\cref{sec:poisson}]
    is another variant, based on the Eisenstein series but aiming for greater small-$t$ stability.
    \item \emph{Wrońskian on the real line} [\cref{sec:spacelike,sec:c1c2}]
    allows for the computation of $\bar E_{5,6}(4;t)$ for $t\in(-\infty,16)$, based on some underlying implementation of $E_1(2;t)$.
    In practice, we only use this method for the precomputation at $t=t_\star=-2$, i.e., \cref{eq:star-values}.
    \item \emph{Contour-integrated Wrońskian} [\cref{sec:timelike,sec:practical}]
    extends the previous method to all $t$ and is numerically faster, given that the integrals at $t_\star$ have been precomputed.
    \item \emph{Series expansion} (\cref{app:t0-series}) is the most dependable method for $t$ close to zero.
    Its radius of convergence is $16$ for $E_{1,2,3}(2;t)$ and $4$ for $\bar E_{4,5,6}(4;t)$.
    It is trivial to obtain very high-order series expansions from the differential equations, and for high-precision applications, we have settled on a 32nd-order one.
    \item \emph{\texttt{pySecDec}} \psdrefs\ is used as our main benchmark.
    It allows for the computation of all master integrals at all $t$, although for $\Re(t)$ above threshold it is typically only dependable for $\Im(t)\geq 0$.
    \item \emph{\texttt{AMFlow}} \cite{Liu:2022chg} is another benchmark that works for all master integrals and all real $t$.
    It is very precise but also very slow and, as we will see below, not always accurate.
    \item \emph{\texttt{FeynTrop}} \cite{Borinsky:2020rqs,Borinsky:2023jdv} is another, faster benchmark, but is restricted to the finite integrals $E_{1,2,3}(2;t)$ [although $\bar E_4(2;t)$ is accessible through \cref{eq:E4-finite}].
\end{itemize}
\noindent Note that our methods are not meant to compete with the general-purpose codes \texttt{pySecDec}, \texttt{AMFlow}, and \texttt{FeynTrop}. Rather, we have implemented them because they deliver more precision, are faster or more stable for the particular integrals considered in this paper.

\subsection{The master integrals as functions of $t$}

\begin{figure}[tp]
    \centering
    \includegraphics{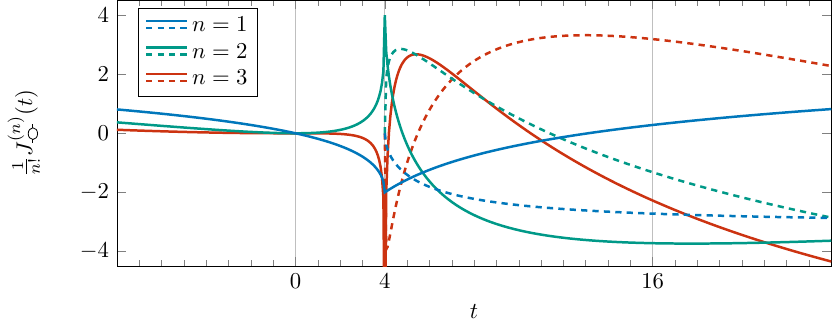}
    \figJbubcaption
    \label{fig:Jbub}
\end{figure}

\begin{figure}[p]
    \hspace{-1cm}
    \includegraphics{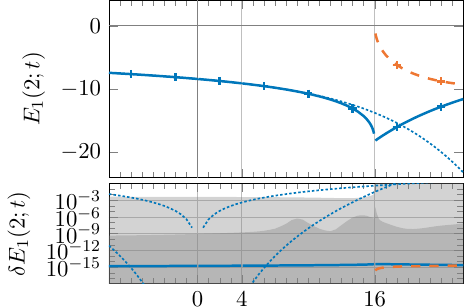}
    \includegraphics{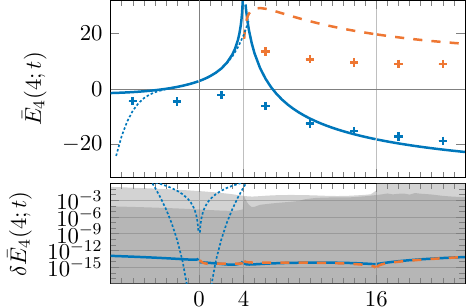}

    \hspace{-1cm}
    \includegraphics{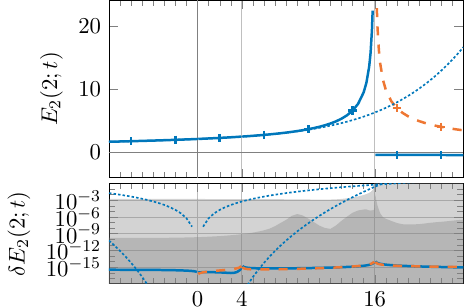}
    \includegraphics{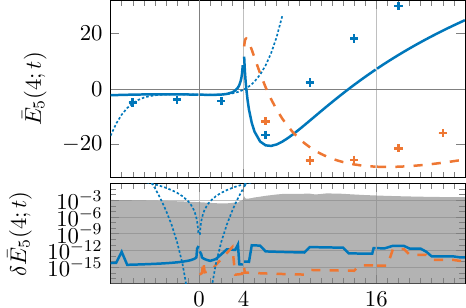}

    \hspace{-1cm}
    \includegraphics{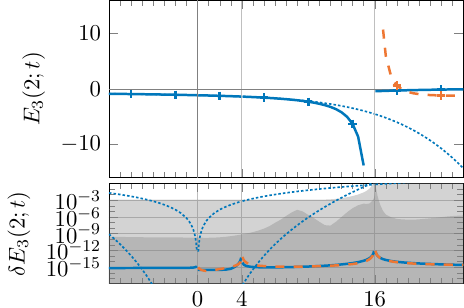}
    \includegraphics{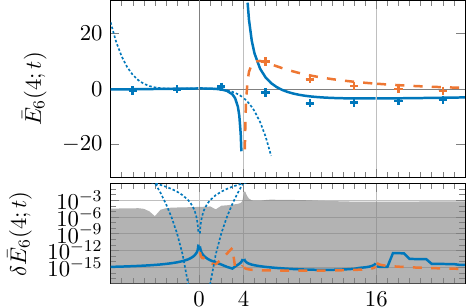}

    \hspace{3cm}$t$\hspace{6.5cm}$t$ 
    
    \figmasterscaption
    \label{fig:masters}
\end{figure}

\Cref{fig:Jbub,fig:masters} show $\Jbub{1,2,3}$ and the six elliptic master integrals along the real $t$ line, including the imaginary part above threshold (taking $\Im t$ small but positive).
Threshold is $t=16$ for $E_{1,2,3}$ and $t=4$ for $\bar E_{4,5,6}$ and $\Jbub{1,2,3}$.
The expected features are visible: a cusp at threshold (or a pole for those master integrals that are derivatives of others), after which the imaginary part develops, indicating a branch cut.  
$\bar E_5$ in particular has a peculiar shape: nearly constant below threshold, and nearly linear for large $t$.
As shown by the error bands, the precision of our method far exceeds that of the benchmarks; furthermore, \texttt{AMFlow} appears unable to accurately compute $\bar E_{4,5,6}(4;t)$.
We return to this point in more detail later in this section.


While our methods work over the entire complex $t$-plane, we have chosen not to show complex counterparts to \cref{fig:masters}.
As can be gleaned from restricting \cref{fig:elliptic} to the outlined region, the qualitative appearance of the master integrals does not differ much from that of a logarithm (possibly times a rational function with zeroes and poles only on the set $\{0,4,16,\infty\}$), so such figures would not look particularly interesting.

\begin{figure}[tp]
    \hspace{-1cm}
    \includegraphics{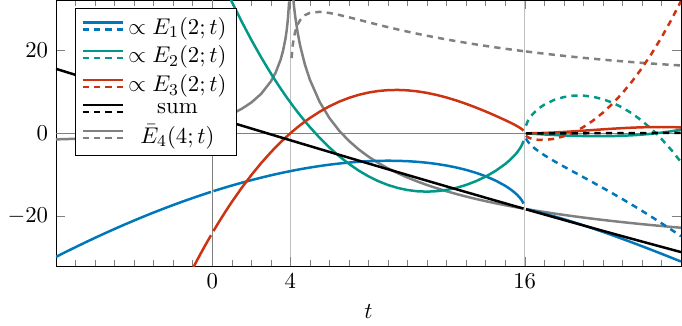}
    \includegraphics{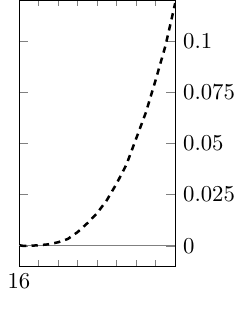}
    \figEfourcaption
    \label{fig:E4}
\end{figure}

An unexpected feature (or rather lack thereof) in $\bar E_{4,5,6}(4;t)$ is that there is no additional cusp at $t=16$, even though they possess the corresponding four-particle cut.
At the amplitude level, this is expected from the phase-space suppression of the four-particle state, but it is not obvious that this should be reflected also at the master integral level; after all, several master integrals possess poles that are not present in the amplitude.
While the intricate nature of $\bar E_5$ and $\bar E_6$ obscures the origin of this,
the case of $\bar E_4$ is easier to study, as is done in \cref{fig:E4}.
There, we see that the contributions from $E_{1,2,3}$ cancel at $t=16$, giving a gently opening cut.

\subsection{Internal comparison of our methods}

\begin{figure}[tp]
    \includegraphics{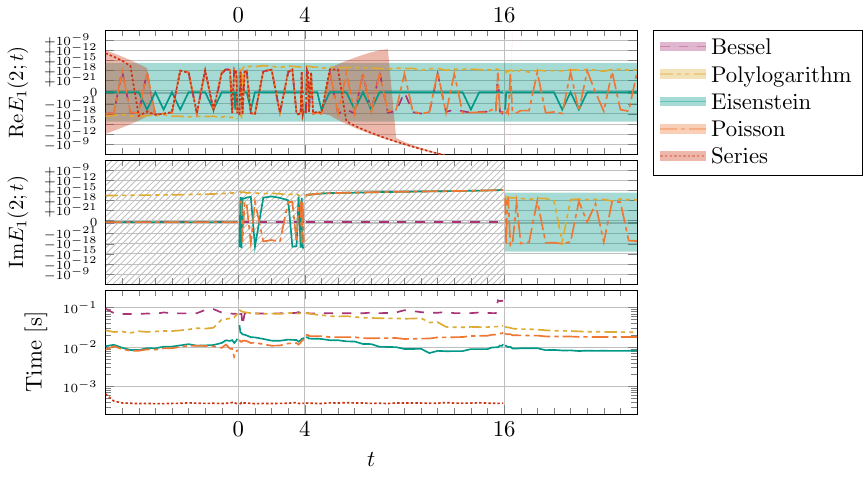}
    \figEonecaption
    \label{fig:E1}
\end{figure}
    
Our implementation makes extensive use of the \texttt{mpmath} library~\cite{mpmath}, which permits arbitrary floating-point working precision.
The numerical integration and summation routines used claim the ability to double their precision by doubling the number of integrand or summand evaluations. 
Even when nested (as is the case for $\IE{n}$), this still implies that the computational complexity of our implementation is polynomial in the number of digits of precision required (quadratic for $\bar E_{5,6}(4;t)$ and linear for the other integrals, compounded with the cost of basic arithmetic scales with the working precision).
However, we have not subjected this scaling to any rigorous tests.
A reasonable working precision is $128$ bits, although we lower that to $64$ for the plots shown in this section, both for speed and to make the error bands easier to see, and to $32$ for the contour integrals in $\bar E_{5,6}(4;t)$, which are time-consuming but numerically stable.
Using mixed precision---low for integrals, medium for functions evaluated at $\beta$, and high for the fixed values at $\beta_\star$, i.e., \cref{eq:star-values}---provides a large speed boost at a moderate cost in precision.

Since all integrals hinge on our implementation of $E_1$, let us study the latter in more detail.
\Cref{fig:E1} compares the five different methods we have to evaluate it: Bessel integrals, the three variants of the elliptic polylogarithm method, and the series expansion.
It is immediately clear that, within their respective domains of validity, all methods are compatible within the error band of the Eisenstein variant.
The exception is the imaginary part below threshold, where all methods are consistent with zero within that same error, but none comes with an error that large.
The non-Eisenstein methods are typically not compatible with one another within their own error bands.
For $-4\lesssim t\lesssim 4$, the Poisson and, to a lesser extent, Bessel implementations appear to be even more precise than the Eisenstein one, judging by how closely they match the series expansion there. Still, their extremely small uncertainty estimates consistently fail to account for the small amount by which they do deviate (not visible on the scale of the plot).

This underestimation of errors is part of a wider tendency that plagues our implementation.
The reason is that we derive the uncertainties entirely from the reported truncation error of the numerical integration or summation, while assuming that the integrands and summands are exact to within the arithmetic working precision.
This is a good assumption for the very simple summand of the Eisenstein series, but evidently less so for the special functions involved in the Poisson and Bessel implementations, and even less so for the polylogarithms, whose relatively poor performance is worsened by the presence of a linear combination of imperfect numerical sums.
This issue is difficult to work around.

In the following, we will exclusively use the Eisenstein implementation to compute the master integrals.
This is due to its more conservative error estimates and its speed.
The Poisson variant outperforms it in some cases, but we have chosen to avoid the complication of using a mix of the two.

\stepcounter{footnote}
\footnotetext{\label{fn:hardware}
    All computations in this work were run on a consumer laptop with an AMD Ryzen 7 PRO 8840U and 31 GB of RAM.}

\subsection{Benchmarking against existing codes}

 \begin{figure}[p]
    \hspace{-1cm}
    \includegraphics{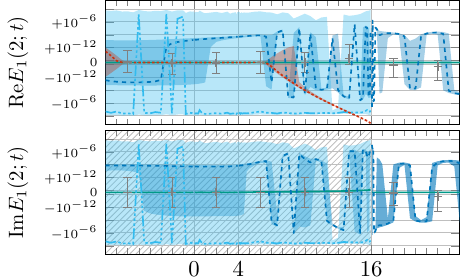}
    \includegraphics{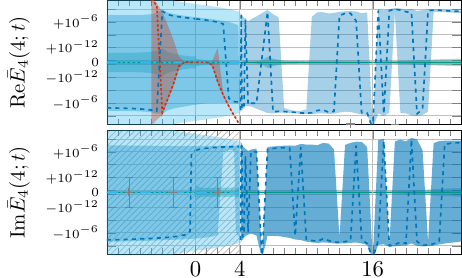}

    \hspace{-1cm}
    \includegraphics{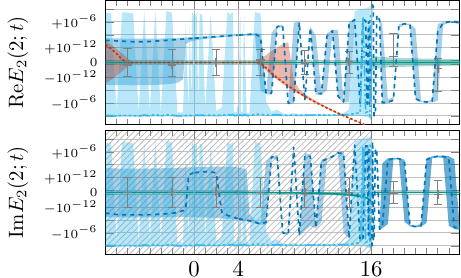}
    \includegraphics{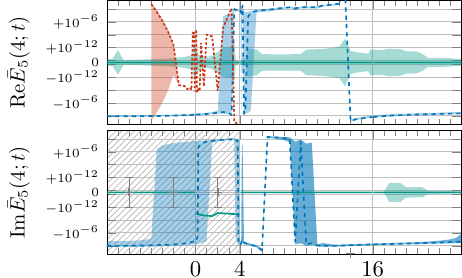}

    \hspace{-1cm}
    \includegraphics{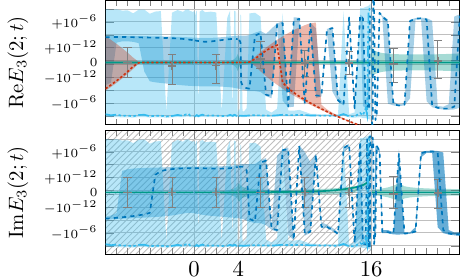}
    \includegraphics{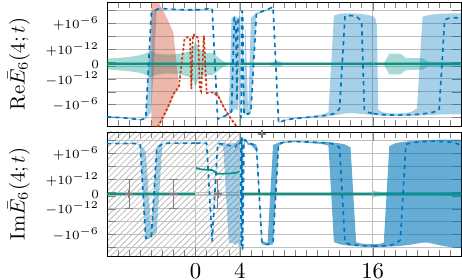}

    \hspace{2.7cm}$t$\hspace{6.8cm}$t$ 

    \figreldiffcaption
    \label{fig:reldiff}
\end{figure}
\afterpage{\clearpage} 

\Cref{fig:reldiff} compares the three benchmark methods against our implementation of the integrals as well as the series expansion.
We use the default settings of \texttt{pySecDec}, $N=10^6$ Monte Carlo samples for \texttt{FeynTrop}, and give \texttt{AMFlow} a precision goal of 12 digits.
The presentation is deliberately similar to \cref{fig:E1}.

As already observed in \cref{fig:masters}, \texttt{AMFlow} fails to compute $\bar E_{4,5,6}(4;t)$, while for $E_{1,2,3}(2;t)$ it is in perfect agreement with our implementation.
As also remarked there, even at this reduced precision, our implementation is much more precise than \texttt{pySecDec}, which in turn is much more precise than \texttt{FeynTrop}.
There is also a tendency for these results to oscillate and deviate from ours by much more than their error estimates would allow.
We now establish that this is due to shortcomings in these general-purpose codes rather than to errors in our implementation.

The most reliable validation of our results comes from the cases where the \emph{exact} value of the integral is known: the imaginary part below threshold, and the series expansion of the real part for small $t$.
The deviation of \texttt{pySecDec} and \texttt{FeynTrop} from these exact values is essentially the same as that from our values. This suggests that our values are indistinguishable from the exact ones at the level of precision of those codes.%
\footnote{
    Fig.~\ref{fig:reldiff} makes \texttt{FeynTrop} appear worse than it is.
    Greater accuracy is possible with more samples, although a factor $100$ increase is needed for each additional digit of precision, as is typical with Monte Carlo methods.
    \texttt{FeynTrop}'s strength lies in that its performance scales well with the number of loops, not in its precision for ``low-loop'' integrals.
    Likewise, $\bar E_{4,5,6}(4;t)$ are quite different from what \texttt{AMFlow} has previously been tested on: it excels on partly massless two-loop integrals with many external legs, which present a wholly different kind of challenge than our fully massive three-loop integrals with a single scale.
    Finally, before we surpassed \texttt{pySecDec}'s accuracy for this small number of integrals, reproducing its output was the standard by which we determined whether our calculations were correct.}

However, our values are clearly not identical to the exact ones.
For $E_{1,2,3}(2;t)$, deviations from $\Im E_i(2;t)=0$ are limited to $\O(10^{-15})$ for $t$ just below $16$.
$\Im\bar E_4(4;t)$ is compatible with zero within errors, but $\bar E_{5,6}(4;t)$ fare worse.
The behavior of the normalized series expansion around $t=0$ indicates that our values oscillate around the true ones with an amplitude approaching $\O(10^{-9})$. Similar deviations from $\Im\bar E_i(4;t)=0$ for $t\in(0,4)$ are also present.
These effects shrink commensurately with increasing working precision, so this is a matter of underestimated uncertainties, similar to those mentioned in the context of \cref{fig:E1}, but more severe.
In fact, we do not fully propagate our errors here: we have turned off the $\delta\int f(\xi)\d\xi\approx\int\delta f(\xi)\d\xi$ estimation mentioned in \cref{sec:practical}, since it takes almost as long to compute as the integral itself and does not increase the error enough to account for the discrepancy.
Better error propagation is unlikely to solve this issue, since we do not fully know the original error of the underlying functions.
Thus, it is more pragmatic to either leave the underestimated errors as they are---after all, \texttt{pySecDec} does quite well while regularly underestimating its uncertainty by at least an order of magnitude---or to add a conservative blanket uncertainty of about $10^{-10}$ (or correspondingly smaller at higher working precision).
In any case, we easily reach a precision comfortably greater than what is practically needed, even more so if the working precision is $128$ bits.

\begin{figure}[tp]
    \hspace{-1cm}
    \includegraphics{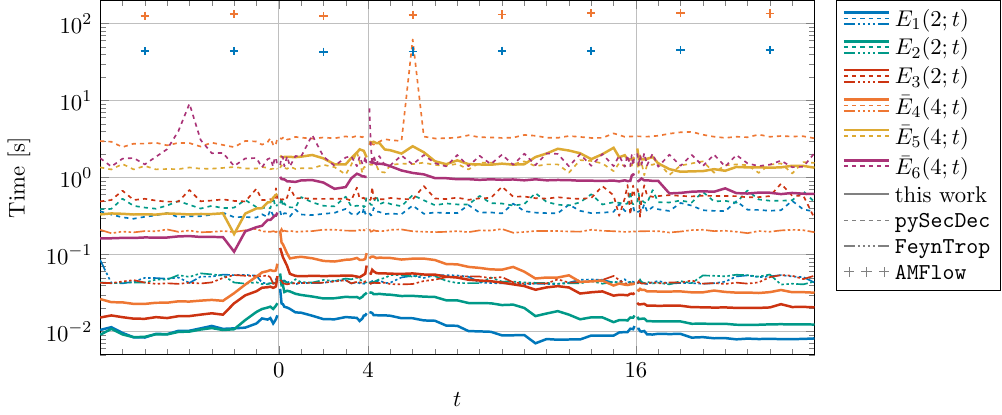}
    \figtimecaption
    \label{fig:time}
\end{figure}

Lastly, let us look at evaluation speed.
Applications such as moments and dispersive integrals require sufficiently many HVP evaluations that master integrals that demand nontrivial computer time should be avoided.
In \cref{fig:time}, we see that our implementation of $E_{1,2,3}$ and $\bar E_4$ is quite fast, comparable to $10^6$-sample \texttt{FeynTrop} and much faster than \texttt{pySecDec}, which struggles with $\bar E_4$ in particular.
On the other hand, our implementation of $\bar E_{5,6}$ is comparable to \texttt{pySecDec}'s speed.
We are much faster in the spacelike region, where the contour integrals can be performed with real numbers, and $\bar E_6$ is about twice as fast as $\bar E_5$ due to $G_{16}=0$ in \cref{eq:Gdef}.

Our implementation is single-threaded, whereas \texttt{pySecDec} and \texttt{FeynTrop} parallelize efficiently.
The data shown here was collected on a 16-thread processor,\hardwarefn
so our single-threaded performance is an order of magnitude better than what \cref{fig:time} would suggest.
On the other hand, we are unable to efficiently scale to larger computers, except by computing different parts of the $t$ domain in parallel.

\subsection{Numerical evaluation of the HVP}

\begin{figure}[tp]
    \hspace{-1cm}
    \includegraphics{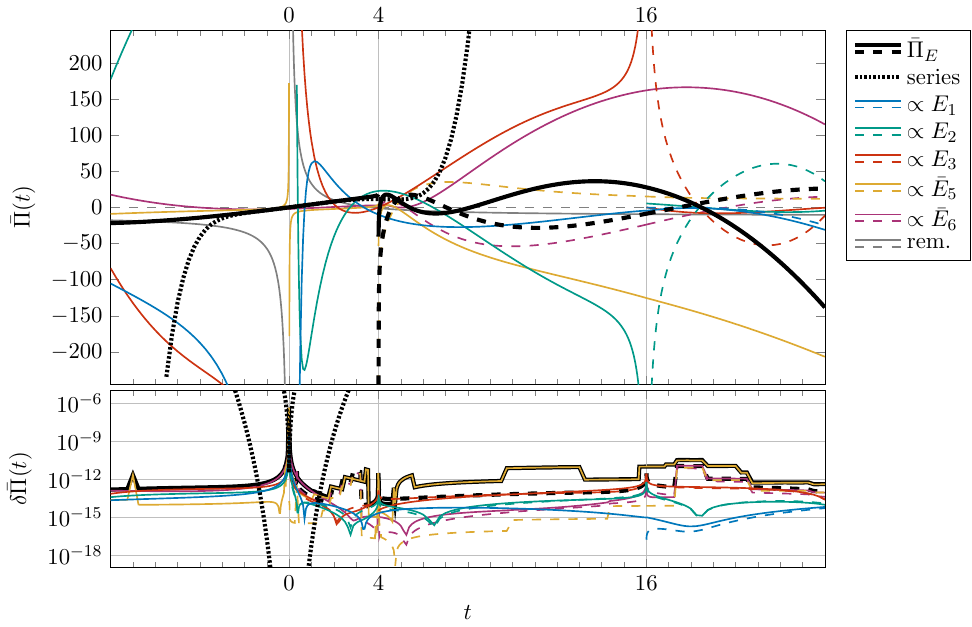}
    \figPiEcaption
    \label{fig:PiE}
\end{figure}
\begin{figure}[tp]
    \hspace{-1cm}
    \includegraphics{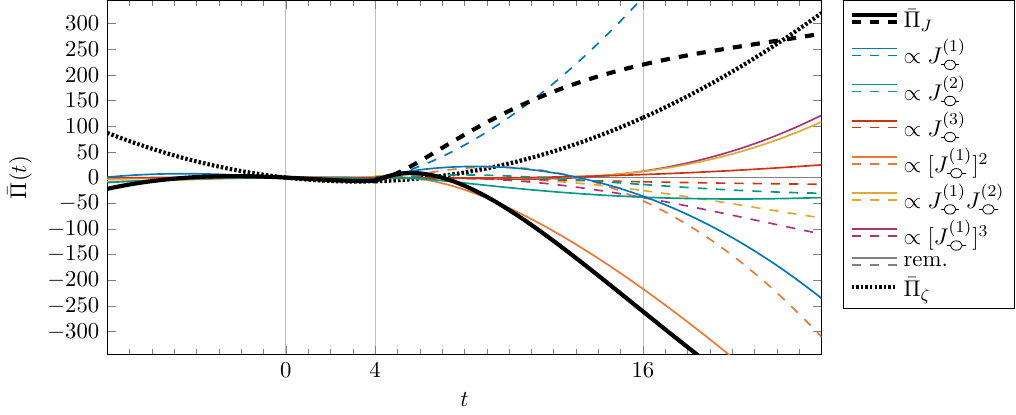}
    \figPiJcaption
    \label{fig:PiJ}
\end{figure}

The full result of \mainref\ depends on several low-energy constants (LEC), the determination of which is beyond the scope of this paper.
However, there are three LEC-less parts of the HVP, namely~\mainref[eq.~(5.5-7)]
{
\allowdisplaybreaks
\begin{align}
    \bar\Pi_E(t)
        &= \bigg[\frac{107 t^3}{46656} - \frac{233 t^2}{3888} - \frac{47 t}{243} + \frac{17327}{1458} - \frac{20813}{486 t} + \frac{1952}{81 t^2}\bigg] E_1(2;t) \notag\\
        &\,- \bigg[\frac{143 t^4}{46656} - \frac{257 t^3}{1944} + \frac{2209 t^2}{1944} + \frac{32287 t}{2916} - \frac{32126}{243} + \frac{72214}{243 t} - \frac{7808}{81 t^2}\bigg] E_2(2;t)\notag\\
        &\,- \bigg[\frac{143 t^4}{23328} - \frac{166 t^3}{729} + \frac{316 t^2}{243} + \frac{29413 t}{1458} - \frac{99722}{729} + \frac{5216}{27 t}\bigg] E_3(2;t) \notag\\
        &\,- \bigg[\frac{t}{3} - \frac{13}{9} - \frac{2}{3t}\bigg] \bar E_5(4;t)
        + \bigg[\frac{t^3}{54} - \frac{19 t^2}{27} + \frac{143 t}{27} - \frac{100}{9}\bigg] \bar E_6(4;t)\notag\\
        &+ \bigg[\frac{1216}{27}+\frac{2 \zeta(3)}{9}-\frac{\pi^2}{18}\bigg]\frac{1}{t}
        - \bigg[\frac{26885}{972}-\frac{2131 \zeta(3)}{108}+\frac{91 \pi^2}{108}\bigg]\,,
    \label{eq:PiE}\\
    \bar\Pi_J(t)
        &= -\bigg[\frac{t^2}{648} + \frac{t}{324} - \frac{1}{108} - \frac{1}{9t}\bigg]\Jbub3(t)
        - \bigg[\frac{t^2}{54} - \frac{8t}{27} + \frac{4}{3} - \frac{2}{3t}\bigg]\Jbub2(t)\,\Jbub1(t)\notag\\
        &- \bigg[\frac{t^2}{54} - \frac{2t}{9} + \frac{8}{9} - \frac{32}{27t}\bigg]\big[\Jbub1(t)\big]^3
        - \bigg[\frac{5 t^2}{216} - \frac{103 t}{108} + \frac{151}{36} + \frac{1}{3 t}\bigg]\Jbub2(t)\notag\\
        &- \bigg[\frac{(1+\Lpi)t^2}{27} - \frac{(63+4\Lpi)t}{54} + \frac{275-64\Lpi}{54} + \frac{10 + 96\Lpi}{27t}\bigg]\big[\Jbub1(t)]^2\notag\\
        &+ \bigg[\frac{(97 -  9\pi^2)t^2}{3888} - \frac{(2285 + 9\pi^2)t}{1944} + \frac{407}{81}+\frac{\pi^2}{72} - \frac{4-\pi^2}{6t}\bigg] \Jbub1(t) \notag\\
        &- \bigg[\frac{t^2 \Lpi}{27} + \frac{(23 + 8\Lpi)t}{27} - \frac{92 + 48\Lpi}{27}\bigg]\Lpi\Jbub1(t)
        - \frac{4-\pi^2}{36}\,,
    \label{eq:PiJ}\\
    \bar\Pi_\zeta(t)
        &= - \frac{t \zeta(3)}{9}
        - \frac{t^2 + 18t}{81} \Lpi^3
        - \frac{20t}{27} \Lpi^2 
        - \bigg[\frac{17t^2}{432} - \frac{845 t}{648}\bigg]\Lpi\notag\\
        &- \bigg[\frac{t^2}{216} - \frac{5 t}{27}\bigg]\pi^2
        + \frac{20245 t^2}{139968} - \frac{55511 t}{23328}\,,
    \label{eq:Piz}
\end{align}}
\noindent where $\Lpi\coloneq2\log(\Mpi/770~\text{MeV})$.
We plot these in \cref{fig:PiE,fig:PiJ}.
As expected, the uncertainty of $\bar\Pi_E$ is entirely dominated by $\bar E_{5,6}$, as is the imaginary part; in fact, the contributions of $E_{1,2,3}$ very nearly cancel in the vicinity of their threshold---for instance, the sum of their imaginary parts is only about $3.6$ even at the largest $t$ shown in \cref{fig:PiE}---meaning that the four-particle cut opens very slowly, and in a similar fashion to the detailed case study in \cref{fig:E4}.

\section{Conclusions and outlook}\label{sec:conclusion}

With this work, we have demonstrated that the formalism of elliptic polylogarithms and Eisenstein series provides a framework for the precise evaluation of the master integrals that enter the three-loop contribution to the hadronic vacuum polarization in chiral perturbation theory. 
We have detailed how to control the analytic structure of the amplitude in the complex $t=q^2/\Mpi^2$ plane, as well as its numerical implementation. 
This reduces the theoretical uncertainty associated with evaluating the Feynman integrals in the master integral basis, making the LECs the main source of uncertainty. 

From a phenomenological perspective, the present numerical implementation has several practical applications. For low-$q^2$ regions, the three-loop analytic form shows that simple polynomials cannot accurately describe hadronic vacuum polarization and require elliptic integrals.
Equipped with this numerical implementation, one can evaluate the four-pion production cross-section $e^+e^-\to 4\pi$, 
where the new LECs do not contribute, so everything needed is known and under good numerical control. 

The approach used in this work combined methods for numerically evaluating the elliptic master integrals. We implemented analytic expressions in terms of elliptic polylogarithms using the formalism of~\cite{Bloch:2014qca} for the finite part of master integrals $\bar E_1(4;t)$, $\bar E_2(4;t)$, and $\bar E_3(4;t)$, and used numerical integration methods for the finite piece of master integrals $\bar E_5(4;t)$ and $\bar E_4(4,t)$.
Elliptic polylogarithms can be represented as iterated elliptic integrals~\cite{Broedel:2017kkb,Broedel:2019hyg,Broedel:2019kmn}. It would be interesting to pursue this formulation as it would provide a more uniform treatment of all the elliptic master integrals appearing in this work, and compare the efficiency of the algorithms for the numerical evaluation of such integrals given in~\cite{Duhr:2019rrs}.

It should be noted that the elliptic nature of the integrals in this work is essentially contained in the evaluation of $E_1(2;t)$ in \cref{sec:E123}, from which $E_2(2;t)$, $E_3(2;t)$ and $\bar E_4(4;t)$ follow as linear combinations of derivatives.
We do not attempt to give similar closed-form expressions for $\bar E_5(4;t)$ or $\bar E_6(4;t)$, instead obtaining them in \cref{sec:E5E6} as extensions of $E_1(2;t)$ using a set of challenging but conceptually simple one-dimensional integrals. 
The same strategy is likely to be viable in other families of integrals, where the most well-behaved member is amenable to analytic treatment and the others can be derived as relatively simple numerical extensions thereof.

\subsection*{Acknowledgements}
MS would like to thank Kálmán Szabo and Volodymyr Biloshytskyi for interesting discussions about alternative approaches to $\bar E_5$.
PV would like to thank Stéphane Lafourcade for some advice on  Python coding.
\texttt{SageMath}~\cite{sagemath}, and \texttt{GNUplot}~\cite{gnuplot} were extensively used in the exploratory stages of the calculations, and for generating some elements of the figures.

The work of LL, AL, and MS was funded in part by the French government under the France 2030 investment plan, as part of the ``Initiative d'Excellence d'Aix-Marseille Université -- A\kern-2pt*MIDEX'' under grant AMX-22-RE-AB-052, and by the Agence Nationale de la Recherche (ANR) under grant HVP4NewPhys (ANR-22-CE31-0011).
The work of PV was funded by the ANR under the grant Observables (ANR-24-CE31-7996)

\appendix
\section{Series expansions around $t=0$}\label{app:t0-series}

As given in \mainref[eq.~(A.20)], the master integrals have the following series expansions around $t=0$:
\begin{subequations}\label{eq:elliptics-t0}
    \begin{align}
        E_1(2;t) &= -7\zeta(3)+\frac{6-7\zeta(3)}{16}t+\frac{54-49\zeta(3)}{1024}
     t^2 + \O(t^3),\\
        E_2(2;t) &= \frac{7 \zeta(3)}{4}+\frac{7 \zeta(3)-6}{32} t +\frac{147 \zeta(3)-162}{4096} t^{2}+ \O(t^3)\,,\\
        E_3(2;t) &= \frac{6-35 \zeta(3)}{32}+\frac{102-105 \zeta(3)}{512} t+\frac{458-399 \zeta(3)}{8192} t^{2} + \O(t^3)\,,\\
        \bar E_4(4;t) &= \frac{-20 + 7\pi^2 - 12\zeta(3)}{12} + \frac{91 + 3\pi^2 - 42\zeta(3)}{48}t + \O(t^2)\,,\\
        \bar E_5(4;t) &= \frac{1}{3} + \frac{\pi^{2}}{12} - \frac{8 \zeta(3)}{3} +\frac{34+8\pi^2-105\zeta(3)}{192} t + \O(t^2)\,,\\
        \bar E_6(4;t) &= -\frac{1}{4}-\frac{\pi^{2}}{16}+\frac{7 \zeta(3)}{8}+\frac{21\zeta(3) - \pi^2 - 15}{96}  t+\O(t^2)\,.
    \end{align}
\end{subequations}
These series can be extended easily by using the differential equations discussed in the main text.%
\footnote{
    Much higher-order versions of the series may be found tabulated in the attached code~\cite{hvpnumerics} in the file \verb|HVPpy/series_expansion.py|, which is generated to $\O(t^n)$ by running \verb|scripts/series.sh |$n$.}
For instance, the series expansion of $E_1(2;t)$ near zero 
\begin{equation}
    E_1(2;t)=\sum_{n=0}^N a_n t^n +\O(t^{N+1})
\end{equation}
and the differential equation \cref{eq:diffeq-E1} give the recurrence relation for the coefficients $a(n)$
\begin{equation}
( n +1)^3 a (n )-2 (2 n +3) (5 n^{2}+15 n +12)a (n +1)+64(n +2)^3 a (n +2)=0\,.
\end{equation}
This recurrence relation, together with the initial conditions (originally derived in \rcite{Bailey:2008ib})
\begin{equation}
    a(0)=-7\zeta(3), \qquad a(1)={6-7\zeta(3)\over 16}
\end{equation}
gives the small-$t$ expansion of $E_1(2;t)$ to any desired order.
One then obtains the  small-$t$ expansions of  $E_2(2;t)$ and $E_3(2;t)$ using their expression as derivatives of $E_1(2;t)$, \cref{eq:E2toE1,eq:E3toE1}.
The expansions of $\bar E_5(4;t)$ and $\bar E_6(4;t)$ are then derived from that of $E_1(2;t)$ using the results of \cref{sec:E5E6}.

\addcontentsline{toc}{section}{References}
\bibliographystyle{JHEP}
\renewcommand\raggedright{}
\bibliography{references}

\end{document}